\documentclass[aps,prd,reprint,onecolumn,notitlepage,nofootinbib]{revtex4-1}
\usepackage{graphicx}
\usepackage{multirow}
\usepackage{amsmath}
\usepackage{bm}
\usepackage{slashed}
\usepackage{epsfig}
\usepackage{amsfonts}
\usepackage{color}
\usepackage{epstopdf}
\usepackage{enumitem}
\allowdisplaybreaks

\newcommand{\ud}{\mathrm{d}}
\newcommand{\uv}{\mathrm{UV}}
\newcommand{\ir}{\mathrm{IR}}

\newcommand{\gev}{\mathrm{~GeV}}

\newcommand{\msbar}{\overline{\mathrm{MS}}}

\newcommand{\state}[4]{{^{#1}\hspace{-0.6mm}#2_{#3}^{[#4]}}}

\newcommand\CSaSz{\state{1}{S}{0}{1}}
\newcommand\CSaPa{\state{1}{P}{1}{1}}
\newcommand\CScSa{\state{3}{S}{1}{1}}

\newcommand\COaSz{\state{1}{S}{0}{8}}

\newcommand\mo{{\mathcal O}}

\newcommand{\LDMEn}[1]{\langle\mo^{#1}_n\rangle}

\newcommand\mohn{\LDMEn{H}}

\def\gev{\mathrm{~GeV}}

\def\co{{\cal O}}

\begin{document}

\title{Semi-analytical calculation of gluon fragmentation into ${^{1}\hspace{-0.6mm}S_{0}^{[1,8]}}$ quarkonia at next-to-leading order} 

\author{Peng Zhang$^{1}$}
\author{Chen-Yu Wang$^{1}$}
\author{Xiao Liu$^{1}$}
\author{Yan-Qing Ma$^{1,2,3}$}
\author{Ce Meng$^{1}$}
\author{Kuang-Ta Chao$^{1,2,3}$}
\affiliation{
$^{1}$School of Physics and State Key Laboratory of Nuclear Physics and
Technology, Peking University, Beijing 100871, China\\
$^{2}$Center for High Energy Physics,
Peking University, Beijing 100871, China\\
$^{3}$Collaborative Innovation Center of Quantum Matter,
Beijing 100871, China
}%
\date{\today}

\begin{abstract}
We calculate the NLO corrections for the gluon fragmentation functions to a heavy quark-antiquark pair in $\CSaSz$ or $\COaSz$ state within NRQCD factorization.
We use integration-by-parts reduction to reduce the original expression to simpler master integrals (MIs), and then set up differential equations for these MIs. After calculating the boundary conditions, MIs can be obtained by solving the differential equations numerically. Our results are expressed in terms of asymptotic expansions at singular points of $z$ (light-cone momentum fraction carried by the quark-antiquark pair), which can not only give FFs results with very high precision at any value of $z$, but also provide fully analytical structure at these singularities. We find that the NLO corrections are significant, with K-factors larger than 2 in most regions. The NLO corrections may have important impact on heavy quarkonia (e.g. $\eta_c$ and $J/\psi$) production at the LHC.
\end{abstract}


\maketitle

\section{Introduction}

Study of heavy quarkonium production is important to understand both perturbative and nonperturbative physics in QCD.
Currently, the most widely used theory for quarkonium production is the nonrelativistic QCD (NRQCD) factorization \cite{Bodwin:1994jh}. Although many important processes have been calculated to next-to-leading order in $\alpha_s$ expansion \cite{Kramer:1995nb,Klasen:2004tz,Zhang:2005cha,Zhang:2006ay,Ma:2008gq,Gong:2009kp,Zhang:2009ym,Campbell:2007ws,Gong:2008ft,Butenschoen:2009zy,Ma:2010vd,Ma:2010yw,Butenschoen:2010rq,Butenschoen:2011ks,Butenschoen:2012px,Chao:2012iv,Gong:2012ug,Wang:2012is,Gong:2013qka,Han:2014kxa,Butenschoen:2014dra,Han:2014jya,Zhang:2014ybe,Bodwin:2015iua}, there are still some notable difficulties in quarkonium production within the NRQCD framework (see, e.g. \cite{Ma:2017xno}). To further explore the quarkonium production mechanism, it may be better to study quarkonium production at high transverse momentum $p_T$ region, where long-distance interactions between quarkonium and initial-state particles are suppressed and thus  factorization is easier to hold.

The inclusive production differential cross section of a specific hadron $H$ at high $p_T$ can be calculated in collinear factorization \cite{Collins:1989gx},
\begin{equation}
	\ud \sigma _{A+B\rightarrow H(p_T)+X} = \sum_i \ud \hat{\sigma} _{A+B \rightarrow i(p_T /z)+X'} \otimes D_{i\to H}(z,\mu) + \co (1/p_T^2) \, ,
\end{equation}
where $i$ sums over all quarks and gluons, $z$ is the light-cone momentum fraction carried by $H$ with respect to the parent parton $i$, and $A$ and $B$ are colliding particles whose effect should be further factorized to partons if they are hadrons.
$d \hat{\sigma} _{A+B \rightarrow i(p_T /z)+X}$ are perturbatively calculable hard parts, while $D_{i\to H}(z,\mu)$ are nonperturbative but universal fragmentation functions (FFs) describing the probability of partons to hardonize to $H$ with momentum fraction $z$. For quarkonium production, $\co (1/p_T^2)$ contributions can be further factorized to double parton FFs \cite{Kang:2011mg,Kang:2014tta,Kang:2014pya,Fleming:2012wy,Fleming:2013qu}.
In both single parton FFs and double parton FFs, there is a collinear factorization scale $\mu$ dependence, and this dependence will be canceled between hard parts and FFs perturbatively order by order, leaving physical differential cross section to be independent of the scale. The evolution of single parton FFs with respect to $\mu$ are controlled by the Dokshitzer-Gribov-Lipatov-Altarelli-Parisi (DGLAP) evolution equation \cite{Gribov:1972ri,Altarelli:1977zs,Dokshitzer:1977sg}, and similar evolution equations for double parton FFs are calculated in \cite{Kang:2014tta}. With these evolution equations, the only unknown information for FFs are their values at a chosen factorization scale $\mu=\mu_f$.

When $\mu_f$ is close to the quarkonium mass $m_H$, it is natural to calculate FFs via NRQCD factorization. For single parton FFs that will be considered in this paper, we have
\begin{equation}
	D_{i\to H}(z,\mu_f) = \sum_n d_{i\to Q\bar{Q}(n)} (z,\mu_f) \langle \bar{\mo}^{H}_n\rangle \, ,
\end{equation}
where $d_{i\to Q\bar{Q}(n)}$ represent the perturbative calculable short-distance coefficients (SDCs) to produce a heavy quark-antiquark pair $Q\bar Q$ with quantum number $n$, and $\langle \bar{\mo}^{H}_n\rangle$ are normalized long-distance matrix elements (LDMEs) \footnote{ $\langle \bar{\mo}^{H}_n\rangle$ can be related to the original definition of NRQCD LDME $\mohn$ \cite{Bodwin:1994jh} by the following rules. They are the same if $n$ is color-octet, and $\langle \bar{\mo}^{H}_n\rangle=\mohn/(2N_c)$ if $n$ is color-singlet.}. The quantum number is usually expressed in spectroscopic notation $n=\state{{2S+1}}{L}{J}{c}$, with $c=1, 8$ respectively for color-singlet state or color-octet state. According to velocity scaling rule \cite{Bodwin:1994jh},  $\langle \bar{\mo}^{H}_n\rangle$ is usually suppressed if $L$ is too large. Therefore, the most important states for phenomenological purpose are $S$-wave and $P$-wave states. Because LDMEs are supposed to be process independent, they can be determined by fitting experimental data, while SDCs need to be calculated perturbatively.

For both $S$-wave and $P$-wave states, all SDCs for single parton FFs are available up to $\alpha_s^2$ \cite{Braaten:2000pc,Ma:1995vi,Braaten:1994kd,Braaten:1993mp,Hao:2009fa,Braaten:1993rw,Jia:2012qx} (see  \cite{Ma:2013yla} for a summary and comparison). However, only a few SDCs have been calculated to $\alpha_s^3$ order, although they are valuable for phenomenological study.  Numerical results for SDCs of $g \rightarrow Q\bar{Q}(\CScSa)+X$ were calculated to LO (order $\alpha_s^3$) in Refs. \cite{Braaten:1993rw,Braaten:1995cj,Bodwin:2003wh,Bodwin:2012xc}, including velocity corrections. Analytical results for this process are only available recently \cite{Zhang:2017xoj} by applying multi-loop techniques developed in the past a few years. Using the same techniques, analytical results for SDCs of $g \rightarrow Q\bar{Q}(\CSaPa)+X$ at LO  (order $\alpha_s^3$) are also obtained \cite{Sun:2018yam}. A more challenging task is the calculation of NLO (order $\alpha_s^3$) SDCs of $g \rightarrow Q\bar{Q}(\CSaSz)+X$, which involves not only tree-level diagrams but also one-loop diagrams. Numerical results for this process have been calculated in Ref.\cite{Artoisenet:2014lpa}. Considering the complicity of the calculation, an independent check by another group is badly needed.

As $\CSaSz$ is the dominant Fock state for $\eta_{c,b}$, the FF $g \rightarrow Q\bar{Q}(\CSaSz)+X$ is important to study  $\eta_{c,b}$ production at high transverse momentum at LHC \cite{Aaij:2014bga}. At the LHC, we have even much more data of $J/\psi$ production at high transverse momentum. Theoretical studies \cite{Ma:2010yw,Chao:2012iv,Bodwin:2014gia,Faccioli:2014cqa} show that $\COaSz$ channel may be crucial to explain the $J/\psi$ data. To calculate $\COaSz$ contribution precisely, we need to calculate the FF $g \rightarrow Q\bar{Q}(\COaSz)+X$ to at least NLO.

In this paper, we aim to calculate NLO SDCs of FFs of $g \rightarrow Q\bar{Q}(\CSaSz)+X$ and $g \rightarrow Q\bar{Q}(\COaSz)+X$ to high precision using similar methods in our previous paper \cite{Zhang:2017xoj}. With sufficient numerical precision, analytical results can in principle be extracted by using PSLQ algorithm. The rest of the paper is organized as following. In Sec. \ref{sec:lo}, we first introduce the definition of SDCs of quarkonium FFs, including projection operators and Feynman rules related to gauge link, and then give the LO results. NLO corrections include real emission Feynman diagrams and one-loop Feynman diagrams, the calculation of them will presented in Sec. \ref{sec:real} and Sec. \ref{sec:virtual}, respectively. In the calculation, we use integration-by-part (IBP) reduction~\cite{Chetyrkin:1981qh,Laporta:2001dd,Studerus:2009ye,Lee:2012cn,Smirnov:2014hma} to express both real contributions and virtual contributions in terms of linear combination of a small set of simpler integrals, which are usually called master integrals (MIs). 
High precision MIs can be obtained easily by solving differential equations of MIs numerically.
Renormalization will be presented in Sec. \ref{sec:renorm}. After renormalization, the obtained SDCs are free of ultraviolet (UV) and infrared (IR) divergences. Final results and discussions will be given in Sec. \ref{sec:summary}. We find that our results for $g \rightarrow Q\bar{Q}(\CSaSz)+X$ seem to be different from that calculated in Ref.\cite{Artoisenet:2014lpa}, while our results for $g \rightarrow Q\bar{Q}(\COaSz)+X$ are new. Finally, high precision results and some technical details will be given in Appendixes.

\section{Calculation of LO FFs} \label{sec:lo}

\subsection{Definitions}
The definition of FF from a gluon to a hadron (quarkonium) is given by Collins-Soper \cite{Collins:1981uw},
\begin{align}\label{eq:defFF}
\begin{split}
    D_{g \rightarrow H}(z,\mu_0)=
    & \frac{-g_{\mu\nu}z^{D-3}}{2 \pi P_c^{+}(N_{c}^{2}-1)(D-2)} \int_{-\infty}^{+\infty}\mathrm{d}x^{-} e^{-i z P_c^{+} x^{-}} \\
    & \times \langle 0 | G_{c}^{+\mu}(0) \mathcal{E}^{\dag}(0,0,\boldsymbol{0}_{\perp})_{cb} \mathcal{P}_{H(P)} \mathcal{E}(0,x^{-},\boldsymbol{0}_{\perp})_{ba} G_{a}^{+\nu}(0,x^{-},\boldsymbol{0}_{\perp}) | 0 \rangle \, ,
\end{split}
\end{align}
where $G^{\mu\nu}$ is the gluon field-strength operator, $P$ and $P_c$ are respectively the momenta of the produced hadron $H$ and initial fragmenting gluon $g$, and $z=P^+/P_c^+$ is the ratio of momenta along the ``+'' direction.
It is convenient to choose the frame in which the hadron has zero transverse momentum, $P = (z P_c^+, m_H^2/(2 z P_c^+),\boldsymbol{0}_{\perp})$, with $P^2=2P^+P^-=m_H^2$.
The projection operator $\mathcal{P}_{H(P)}$ is defined by
\begin{equation}\label{eq:projectH}
\mathcal{P}_{H(P)} = \sum_X |H(P)+X \rangle \langle H(P)+X|\,,
\end{equation}
where $X$ sums over all unobserved particles.
The gauge link $\mathcal{E}(x^{-})$ is an eikonal operator that involves a path-ordered exponential of gluon field operators along a light-like path,
\begin{equation}
  \mathcal{E}(0,x^{-},\boldsymbol{0}_{\perp})_{ba}= \mathrm{P} \, \text{exp} \left[+i g_s \int_{x^{-}}^{\infty}\mathrm{d}z^{-} A^{+}(0,z^{-},\boldsymbol{0}_{\perp}) \right]_{ba} \, ,
\end{equation}
where $g_s=\sqrt{4 \pi \alpha_s}$ is the QCD coupling constant and $A^{\mu}(x)$ is the matrix-valued gluon field in the adjoint representation: $[A^{\mu}(x)]_{ac} = i f^{abc} A^{\mu}_{b}(x)$.

\begin{figure}[htb!]
\begin{center}
\begin{picture}(250,160)
\put(60,130){\makebox(0,0){\includegraphics[width=100pt]{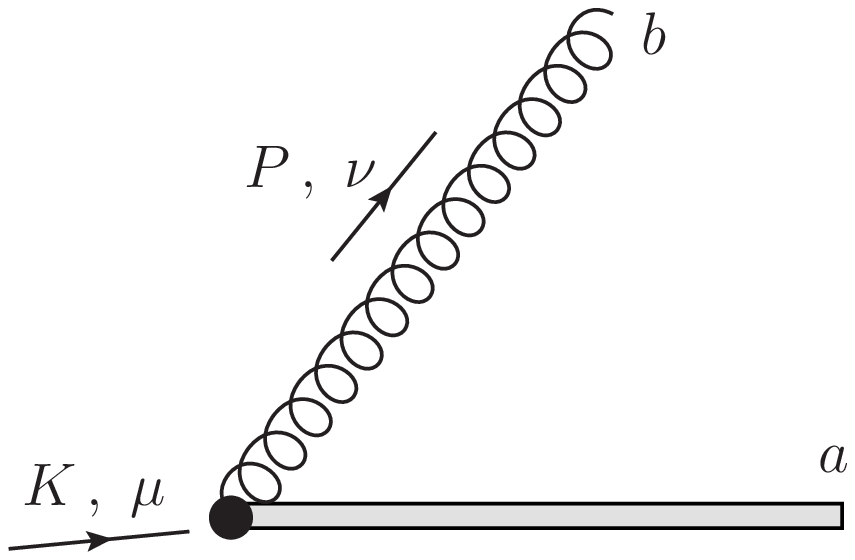}}}
\put(120,130){\makebox(0,0)[l]{${}=-i \delta^{a\,b} \left( g^{\mu\,\nu} - \frac{P^{\mu} \, n^{\nu}}{K\cdot n} \right) $}}
\put(70,70){\makebox(0,0){\includegraphics[width=90pt]{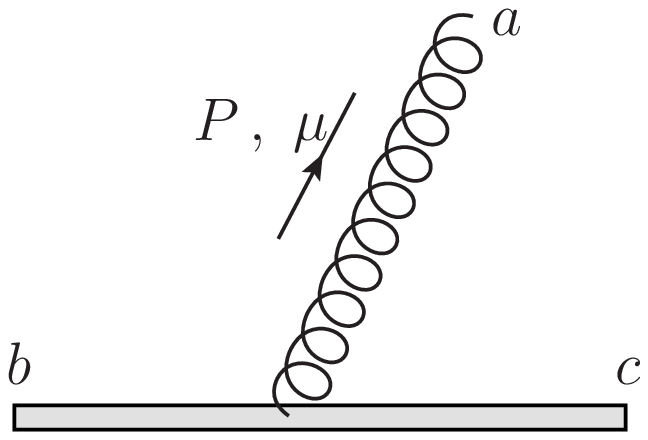}}}
\put(120,65){\makebox(0,0)[l]{${}=g_s f^{a\,b\,c} n^{\mu} $}}
\put(63,5){\makebox(0,0){\includegraphics[width=75pt]{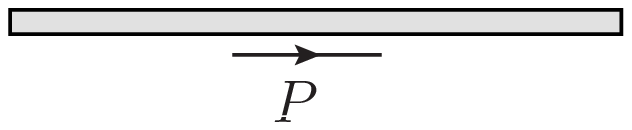}}}
\put(120,10){\makebox(0,0)[l]{${}= i/(P\cdot n +i \varepsilon) $}}
\end{picture}
\end{center}
\caption{Feynman rules related to the gluon gauge link.}
\label{fig:Feynrules}
\end{figure}

From this definition, we can derive Feynman rules related to gauge link, which are showed in Fig.~\ref{fig:Feynrules}, where $n= (0,1^-,\boldsymbol{0}_{\perp}) $, $K$ and $P$ denote momenta, $\mu$ and $\nu$ denote Lorentz indexes, and $a,b$ and $c$ denote color indices.

With these Feynman rules, we can obtain the amplitude of all Feynman diagrams denoted as $\mathcal{M}_{\lambda_{{Q}}\lambda_{\bar{Q}}\lambda_0 \lambda_i}(P,k_{i},m_Q)$, where $\lambda_{{Q}}$ and $\lambda_{\bar{Q}}$ are respectively spins of produced on-shell heavy quark and heavy antiquark, $\lambda_0$ and $\lambda_i$ ($i=1, 2, \dots$) are spins of the initial-state virtual gluon and final-state unobserved light particles, respectively, $k_i$ are the momenta of final-state light particles, and $m_Q$ is the heavy quark mass.
For the processes of gluon fragmenting to S-wave quarkonium, the relative momentum between the $Q\bar{Q}$ pair can be chosen as 0 directly at the lowest order in velocity expansion, and thus it does not appear in the amplitude.
If we project the free $Q\bar{Q}$ pair to specific states $\CSaSz$ or $\COaSz$, we have
\begin{equation}
	\mathcal{M}_{\lambda_0 \lambda_i}(P,k_{i},m_Q)=\text{Tr}\left[ \Gamma_c \Gamma_5 \mathcal{M}_{\lambda_{{Q}}\lambda_{\bar{Q}}\lambda_0 \lambda_i}(P,k_{i},m_Q)  \right]\, ,
\end{equation}
where $\Gamma_c \, ,\Gamma_5$ are the projection operators defined as
\begin{align}
\begin{split}
	&\Gamma_{c=1}
	 = \frac{1} {\sqrt{N_{c}}} \,, \\
	&\Gamma_{c=8}
	 = \frac{\sqrt{2} T^{a}} {\sqrt{N_{c}^{2}-1}} \,, \\
	&\Gamma_{5}
	 = \frac{1} {\sqrt{M} (M/2 + m_Q)}
		(\slashed{P}/2 - m_Q)
		\frac{M - \slashed{P}}{2M}
		\gamma^5
		\frac{M + \slashed{P}}{2M}
		(\slashed{P}/2 - m_Q)	\, ,	
\end{split}
\end{align}
where $P^2=M^2=4m_Q^2$.
By summing over spin and color of initial-state and final-state particles, we get the squared amplitude
\begin{equation}
	\left| \mathcal{M}({P},{k}_{i},m_Q) \right|^2=\overline{\sum} \left| \mathcal{M}_{\lambda_0 \lambda_i}(P,k_{i},m_Q) \right|^2.
\end{equation}
Then the SDCs for gluon fragmenting to spin-singlet S-wave quarkonium can be written as
\begin{equation}\label{eq:sdcdef0}
	d(z) = N_{\mathrm{CS}} \int \mathrm{d} \Phi \, \left| \mathcal{M}(P, k_i, m_Q) \right|^2 \, ,
\end{equation}
where $N_{\mathrm{CS}}=\frac{z^{D-2}}{(N_{c}^{2}-1)(D-2)} $ with $D=4-2\epsilon$ is the space-time dimension, and final-state phase space is defined as
\begin{align}\label{eq:phase}
\begin{split}
  \mathrm{d} \Phi &=
  	\frac {1}{S}
  	\delta \left( z - \frac{P^{+}}{P_{c}^{+}} \right)
    (2\pi)^{D} \delta^{D} \left(
    P_{c} - P - \sum_i k_i \right)
    \frac{\mathrm{d}^{D} P_{c}}{(2\pi)^{D}}
    \prod_{i} \frac{\mathrm{d} k_{i}^{+}}{4\pi k_{i}^{+}}
    \frac{\mathrm{d}^{D-2} k_{i\perp}}{(2\pi)^{D-2}}
    \theta(k_{i}^{+}) \\
    &=
	\frac {P^{+}}{z^{2} S}
      \delta \left( \frac{1-z}{z} P^{+} - \sum_i k_{i}^{+} \right)
      \prod_{i} \frac{\mathrm{d} k_{i}^{+}}{4\pi k_{i}^{+}}
      \frac{\mathrm{d}^{D-2} k_{i\perp}}{(2\pi)^{D-2}}
      \theta(k_{i}^{+}) \,
\end{split}
\end{align}
where $S$ is the symmetry factor for final-state particles.

To be convenient, we extract the dependence on $m_Q$ explicitly by rescaling momenta in the delta function in Eq.~\eqref{eq:phase} by $M$,
\begin{equation} \label{eq:dless}
  \hat{P} = \frac {P}{M} \, , \quad
  \hat{k}_i = \frac {k_i}{M} \, , \quad \hat{m}_Q=\frac{m_Q}{M}=\frac{1}{2} \, .
\end{equation}
Thus the phase space in Eq.~\eqref{eq:phase} changes to
\begin{equation}
	\ud \Phi = M^{n(D-2)} \ud \hat{\Phi} \, ,
\end{equation}
where $n$ is the number of final-state light particles, and $\ud \hat{\Phi}$ is similar to $\ud \Phi$ by changing all momenta to the dimensionless ones.
If we further denote
\begin{equation}
	\hat{\mathcal{M}}_{\lambda_0 \lambda_i}(\hat{P},\hat{k}_{i},m_Q) = M^{n(D-2)/2} \mathcal{M}_{\lambda_0 \lambda_i}(M \hat{P}, M \hat{k}_{i},M \hat{m}_Q) \, ,
\end{equation}
we get a similar relation as that in Eq.~\eqref{eq:sdcdef0},
\begin{equation}
	d(z) = N_{\mathrm{CS}} \int \mathrm{d} \hat{\Phi} \, \left| \hat{\mathcal{M}}(\hat{P},\hat{k}_i, \hat{m}_Q) \right|^2 \,,
\end{equation}
which means that the same SDCs can be obtained by replacing all momenta by their corresponding rescaled ones. In the rest of the paper, we will only use the rescaled momenta, but omitting the `` $\hat{}$ '' for simplicity.

\subsection{LO SDCs}
\begin{figure}[htb!]
 \begin{center}
 \includegraphics[width=0.25\textwidth]{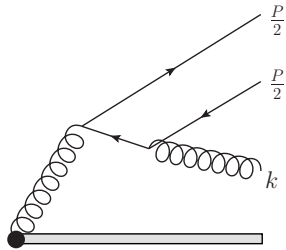}
  \caption{One of the two Feynman diagrams of gluon fragmenting into $\CSaSz$ or $\COaSz$ $Q\bar{Q}$ at LO in $\alpha_s$. Another diagram can be obtained by permuting the heavy quark and anti-quark.  \label{fig:lofd}}
 \end{center}
\end{figure}

The Feynman diagrams of gluon fragmenting into $\CSaSz$ or $\COaSz$ $Q\bar{Q}$ at LO in $\alpha_s$ are showed in Fig. \ref{fig:lofd}.
From the definition above, the calculation of LO SDCs involve integrals of the form
\begin{equation}
	\int \ud \Phi_{\mathrm{Born}} \frac{1}{k\cdot P +a} \, ,
\end{equation}
where $a$ equals $0$ or $1/2$, $k$ is the momentum of the emitted gluon with $k^+ = (1-z) P^+ /z $ and $k^- = k_{\perp}^2 /(2k^+) $, and
\begin{equation} \label{eq:phborn}
	\int \ud \Phi_{\mathrm{Born}} =
		\frac {1}{4\pi z (1-z)}
	     \int \frac{\mathrm{d}^{D-2} k_{\perp}}{(2\pi)^{D-2}} \, .
\end{equation}
These integrals can be performed easily.

Then we get LO SDCs:
\begin{align}
 	d_{\mathrm{LO}}^{[1]} (z) &= \frac{\alpha_s^2}{2(1-\epsilon) N_c m_Q^3 }
 		\left( \frac{\pi \mu_r^2}{m_Q^2} \right)^{\epsilon} d_{\mathrm{LO}} (z) \, , \label{eq:lo1S01}\\
 	d_{\mathrm{LO}}^{[8]} (z) &= \frac{\alpha_s^2 (N_c^2-4)}{4(1-\epsilon) N_c (N_c^2-1) m_Q^3 }
 		\left( \frac{\pi \mu_r^2}{m_Q^2} \right)^{\epsilon} d_{\mathrm{LO}} (z)\,  \label{eq:lo1S08},
\end{align}
where $\mu_r$ is the renormalization scale, $d_{\mathrm{LO}}^{[1]}$ and $d_{\mathrm{LO}}^{[8]}$ respectively denote SDCs of gluon fragmenting into $\CSaSz$ and $\COaSz$ states, and
\begin{equation}
 	d_{\mathrm{LO}} (z) = \Gamma (\epsilon )
 		(2 \epsilon -1) (1-z)^{-2 \epsilon }
 		\left[\left(z (\epsilon^2 -\epsilon +2)-2\right) (1-z)^{\epsilon }+2 (z-1) (z \epsilon -1)\right] \, ,
\end{equation}
with
\begin{equation}\label{eq:lo0}
	d_{\mathrm{LO}}^{(0)} (z) = \lim_{\epsilon \rightarrow 0} d_{\mathrm{LO}} (z) = (3-2 z) z + 2 (1-z) \ln (1-z) \,.
\end{equation}
The color-singlet result and color-octet result are consistent with Refs. \cite{Artoisenet:2014lpa} and \cite{Jia:2012qx} respectively.

\section{Real NLO corrections} \label{sec:real}

\subsection{Reduction to MIs}

\begin{figure}[htb!]
 \begin{center}
 \includegraphics[width=0.6\textwidth]{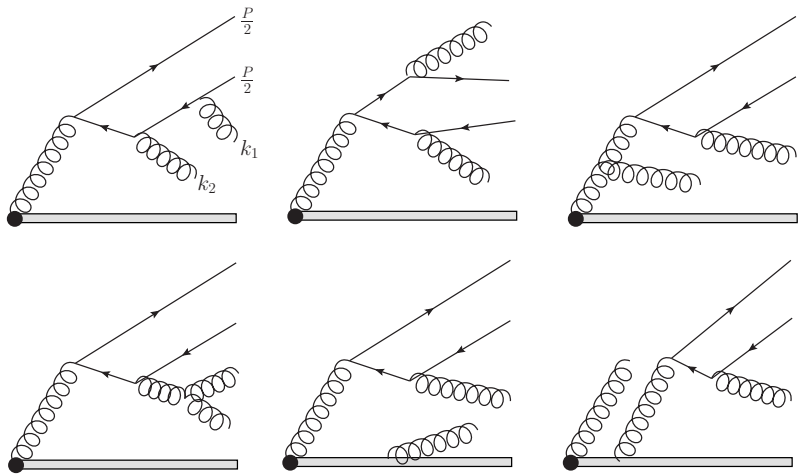}
  \caption{Typical Feynman diagrams for $g\to Q\bar{Q}(\state{1}{S}{0}{1,8})+gg$. The other diagrams can be obtained by permuting the heavy quark and anti-quark or the two emitted gluons.  \label{fig:realggfd}}
 \end{center}
\end{figure}

\begin{figure}[htb!]
 \begin{center}
 \includegraphics[width=0.2\textwidth]{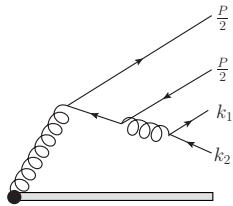}
  \caption{One of the two Feynman diagrams for $g\to Q\bar{Q}(\state{1}{S}{0}{1,8})+q\bar{q}$. Another diagram can be obtained by permuting the heavy quark and anti-quark.  \label{fig:realqqfd}}
 \end{center}
\end{figure}

Real NLO corrections to FFs of $g\to Q\bar{Q}(\state{1}{S}{0}{1,8})+X$ come from Feynman diagrams with two real light particles in the final state, either two gluons or a light quark-antiquark $(q\bar{q})$ pair.
Feynman diagrams with two gluons emission are showed in Fig. \ref{fig:realggfd}, while those with $q\bar{q}$ pair emission are showed in Fig. \ref{fig:realqqfd}.

SDCs can be expressed as linear combinations of integrals of the form
\begin{equation}\label{eq:realint}
	\int \ud \Phi_{\mathrm{real}} \prod_{i} \frac{1}{E_i^{a_i}} =
		\frac{P \cdot n}{2 z^{2}}
		\int \frac{\mathrm{d}^{D} k_1}{(2\pi)^{D-1}}
        \frac{\mathrm{d}^{D} k_2}{(2\pi)^{D-1}}
        \delta _+ (k_1^{2})
        \delta _+ (k_2^{2})
        \delta \left( k_1 \cdot n + k_2 \cdot n - \frac{1-z}{z} P \cdot n \right)
        \prod_{i} \frac{1}{E_i^{a_i}} \, ,
\end{equation}
where $a_i$ are integers, $k_1$ and $k_2$ are momenta of the final-state light particles, the phase space $\ud \Phi_{\mathrm{real}}$ is defined in Eq.~\eqref{eq:phase} with $S=2$, and
\begin{align} \label{eq:edef}
\begin{split}
&	E_1 = k_1 \cdot k_2 \, , \
	E_2 = k_1 \cdot P \, , \
	E_3 = k_2 \cdot P \, , \
	E_4 = 2 \, k_1 \cdot P + 1 \, , \
	E_5 = 2 \, k_2 \cdot P + 1 \, , \ \\
&	E_6 = 2 \, k_1 \cdot k_2 + k_1 \cdot P + k_2 \cdot P \, , \
	E_7 = 2 \, k_1 \cdot k_2 + 2 \, k_1 \cdot P + 2 \, k_2 \cdot P + 1 \, , \ \\
&	E_8 = k_1 \cdot n \, , \
	E_9 = k_1 \cdot n + P \cdot n \, , \
	E_{10} = k_2 \cdot n \, , \
	E_{11} = k_2 \cdot n + P \cdot n \, . \
\end{split}
\end{align}
In Eq.~\eqref{eq:realint}, we safely ignore infinitesimal imaginary parts in denominators because $E_i (i=1,\cdots 11)$ are positive definite and that SDCs are well regularized by dimensional regularization. The later condition implies that only the region where all $E_i (i=1,\cdots 11)$ are not too small can contribute to the phase space integration.  Note that, for $q\bar{q}$ pair emission, although the symmetry factor should be 1, we can also express the SDCs as linear combinations of integrals in Eq.~\eqref{eq:realint}.

To take advantage of multi-loop techniques, we express delta functions by propagator denominators,
\begin{equation} \label{eq:delta}
	(2\pi)\delta (x) =
		\lim_{\eta \rightarrow 0^+}
		\left(
		\frac{i}{x + i \eta}
		+ \frac{-i}{x - i \eta}
		\right) \, .
\end{equation}
We replace the three delta functions in Eq.~\eqref{eq:realint} following the above rule, and denote
\begin{equation}
E_{12} = k_1^2 \, , \
E_{13} = k_2^2 \, , \
E_{14} = k_1 \cdot n +k_2 \cdot n -\frac{1-z}{z} P \cdot n \, .
\end{equation}
Then each phase space integral in Eq.~\eqref{eq:realint} is translated to 8 loop integrals, with either positive or negative infinitesimal imaginary parts in new denominators.

If we forget about infinitesimal imaginary parts in denominators for the moment, we need to deal with loop integrals
\begin{equation}
\int \frac{\mathrm{d}^{D} k_1}{(2\pi)^{D}}
\frac{\mathrm{d}^{D} k_2}{(2\pi)^{D}}
\prod_{i=1}^{14} \frac{1}{E_i^{a_i}} \,
\end{equation}
with integers $a_i$, which can be expressed in terms of corresponding simpler MIs by using IBP reduction~\cite{Chetyrkin:1981qh,Laporta:2001dd,Studerus:2009ye,Lee:2012cn,Smirnov:2014hma}. MIs are also the same kind of integrals, but usually with smaller $a_i$. Note that, we can always choose MIs with powers of $E_{12},E_{13}$ and $E_{14}$ being no larger than 1. For MIs with integrand involving $\frac{1}{E_{12}}$, we can replace the denominator by $\delta _+ (k_1^{2})$ considering the relation Eq.~\eqref{eq:delta}, while for MIs with integrand $E_{12}^{-a_{12}}$ ($a_{12}\leq0$) we can set it to zero. Similar replacement can be done for $E_{13}$ and $E_{14}$. Therefore, all MIs for loop integration are changed back to corresponding MIs for phase space integration defined in Eq.~\eqref{eq:realint}. Once these MIs are also calculated, we obtain final results of real corrections.

In the above procedure, we actually assume that IBP reduction relations are independent of infinitesimal imaginary parts in denominators. This assumption, unfortunately, does not always hold. If one or more integrals cannot be fully regularized by dimensional regularization, one may get wrong final results. In the Appendix \ref{sec:reg}, we will discuss this problem in more details, and then propose a solution. Eventually, the above procedure is justified with a small modification.

\subsection{Calculation of MIs} 
To calculate these MIs, we use differential equations (DEs) method~\cite{Kotikov:1990kg,Bern:1992em,Remiddi:1997ny,Gehrmann:1999as,Henn:2013pwa, Lee:2014ioa,Adams:2017tga,Caffo:2008aw,Czakon:2008zk,Mueller:2015lrx,Lee:2017qql,Liu:2017jxz,Liu:2018dmc}, which has also been used in our previous paper \cite{Zhang:2017xoj} to calculate SDCs of $g \rightarrow Q\bar{Q}(\CScSa)+X$. We get $95$ MIs using the IBP reduction program FIRE5 \cite{Smirnov:2014hma}, without using the symmetry rules. We set up DEs by first differentiating these MIs $I_k(k=1,\dots,95)$ with respect to $z$, and then reducing the resulted integrals to MIs again by using IBP reduction, which results in
\begin{equation}\label{eq:de}
	\frac{\ud \boldsymbol{I} (\epsilon,z)}{\ud z}
	= {A} (\epsilon, z) \boldsymbol{I} (\epsilon,z) \, ,
\end{equation}
where $\boldsymbol{I}$ represents the vector of MIs $I_k$, and ${A}$ is a $95\times95$ matrix whose elements are rational functions of $z$ and $\epsilon$. Having the DEs, we also need boundary conditions of $I_k$ to fully determine these MIs. We choose the boundary at $z\to1$, and calculate the boundary conditions in Appendix \ref{sec:BCreal}.

With boundary conditions, we can solve the DEs to obtain MIs at any value of $z$. One possible choice is to solve the DEs analytically, which can be done by transforming DEs into canonical form (or $\epsilon$-form) \cite{Henn:2013pwa,Lee:2014ioa}. In this way, we successfully express MIs in terms of Goncharov polylogarithms (GPLs)\cite{Goncharov:2001iea}. All obtained GPLs have weights at most three, and they can be expressed in terms of logarithms and classical polylogarithms $\text{Li}_n(z),(n\leq3)$ \cite{Frellesvig:2016ske}. Even though, the obtained analytical expression is too long to present in this paper. Furthermore, for virtual correction, boundary conditions are hard to calculate analytically.

Another choice is to solve DEs numerically, which is a well-studied mathematical problem. 
DEs can help to do asymptotic expansions of MIs around any point $z=z_0$. Because Feynman integrals have Feynman parametric representation, their asymptotic expansions have the form (see e.g. Ref. \cite{Henn:2014qga})
\begin{equation}\label{eq:miexp}
	I_k(z,\epsilon)|_{z_0} = \sum_{s} \sum_{i=0}^{n_{s}} (z-z_0)^{s} \ln^{i}(z-z_0) \sum_{j=0}^{\infty} I_{k}^{ s \, i \, j}(\epsilon) (z-z_0)^j \, ,
\end{equation}
where $s$ is a linear function of $\epsilon$, $n_s$ is an integer determined by $s$, $I_{k}^{ s \, i \,j }(\epsilon)$ are functions of $\epsilon$, and the radius of convergence of the summation over $j$ is usually determined by the nearest singular point. For the special case when $z_0$ is an analytical point, we have $s=n_s=0$. When $z_0$ is a singular point,
different values of $s$ and $i$ correspond to different regions of MIs, which are independent of each other. Therefore, each region satisfies the same DEs as the original MIs, and the DEs can generate recurrence relations to express $I_{k}^{ s \, i \,j }(\epsilon)$ in terms of $I_{k}^{ s \, 0 \,0 }(\epsilon)$ for each fixed $s$, $i$ and $\epsilon$. It implies that, when calculating boundary conditions in Appendix \ref{sec:BCreal}, we only need to calculate $I_{k}^{ s \, 0 \,0 }(\epsilon)$ for each region. In practice, as we are only interested in MIs up to a fixed order in $\epsilon$ expansion, we will do a Laurent expansion of $\epsilon$ in both $(z-z_0)^{s}$ and $I_{k}^{ s \, i \,j }(\epsilon)$.

\begin{figure}[htb!]
	\begin{center}
		\includegraphics[width=0.4\textwidth]{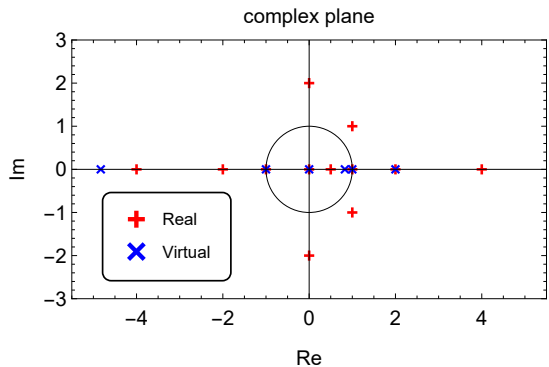}
		\caption{Singularities of DEs of MIs for both $g\to Q\bar{Q} (\CSaSz) +X$ and $g\to Q\bar{Q} (\COaSz) +X$. Plus signs denote singularities encountered in real corrections while  multiplication signs denote singularities encountered in virtual corrections.\label{fig:depole}}
	\end{center}
\end{figure}

As it is clear, singular points play important role in the procedure of solving DEs numerically. There are 12 singular points in the DEs~\eqref{eq:de} for real corrections, which are located at $z=0, 1/2, \pm1, \pm2, \pm 4, \pm 2i, 1\pm i$, as shown in Fig.~\ref{fig:depole}. 
For the interested physical region $0\leq z\leq1$, the only relevant singularities are $z=0,1/2,1$, and all other singularities are far enough from the physical region. Among these three singularities, the point $z=1/2$ is in fact a removable singularity. However, as we will discuss in Appendix \ref{sec:singular}, this singularity determines the radius of convergence  of the asymptotic expansion at $z=0$ and $1$.  We thus estimate values of MIs in regions $0\sim1/4$, $1/4\sim3/4$ and $3/4\sim1$ respectively by the asymptotic expansions of MIs at $z=0,1/2$ and $1$. 
For example, if we want to obtain values in the physical region with precision about $15$ digits, we should calculate the expansion in Eq.~\eqref{eq:miexp} with $j$ to as large as $50$.

\section{Virtual NLO corrections} \label{sec:virtual}
\begin{figure}[htb!]
 \begin{center}
 \includegraphics[width=0.9\textwidth]{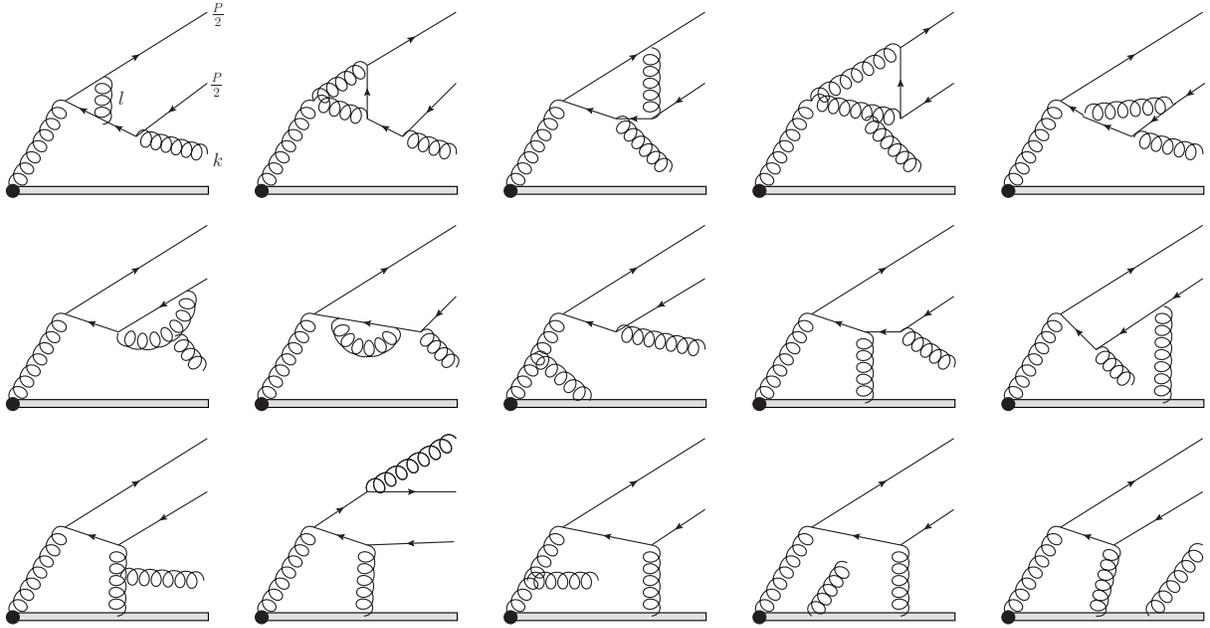}
  \caption{Some typical Feynman diagrams of the virtual NLO correction for gluon fragmenting into $\CSaSz$ or $\COaSz$ $Q\bar{Q}$. The other diagrams are either self-energy diagrams for external legs (including initial virtual gluon), or they can be obtained by permuting the heavy quark and anti-quark.  \label{fig:virtualfd}}
 \end{center}
\end{figure}

Some diagrams that contributed to virtual NLO corrections to FFs of $g\to Q\bar{Q}(\state{1}{S}{0}{1,8})+X$ are shown in Fig. \ref{fig:virtualfd}. The other diagrams are either self-energy diagrams for external legs (including initial virtual gluon), or they can be obtained by permuting the heavy quark and anti-quark.

SDCs of the virtual corrections can be expressed as linear combination of integrals of the form
\begin{equation}\label{eq:virint}
	\int \ud \Phi_{\mathrm{loop}}
	\int \frac{\mathrm{d}^{D} l}{(2\pi)^{D}} \prod_{i} \frac{1}{F_i^{a_i}} =
		\frac{P \cdot n}{z^{2}}
		\int \frac{\mathrm{d}^{D} k}{(2\pi)^{D-1}}
        \frac{\mathrm{d}^{D} l}{(2\pi)^{D}}
        \delta _+ (k^{2})
        \delta \left( k \cdot n - \frac{1-z}{z} P \cdot n \right)
        \prod_{i} \frac{1}{F_i^{a_i}} \, ,
\end{equation}
where $z=P^+/(k^+ + P^+)$, $a_i$ are integers, $k$ is the momentum of the final-state gluon, $l$ is the loop momenta, and
\begin{align} \label{eq:edefloop}
\begin{split}
&	F_1 = k \cdot P \, , \
	F_2 = 2 \, k \cdot P + 1 \, , \
	F_3 = l^2 \, , \
	F_4 = (l+k)^2 \, , \
	F_5 = (l+P)^2 \, , \ \\
&	F_6 = (l+\frac{P}{2})^2 -\frac{1}{4} \, , \
	F_7 = (l-\frac{P}{2})^2 -\frac{1}{4} \, , \
	F_8 = (l+k+\frac{P}{2})^2 -\frac{1}{4} \, , \
	F_9 = (l+k+P)^2  \, , \
	F_{10} = l \cdot n  \, . \
\end{split}
\end{align}
Similar to real corrections, by replacing $\delta$ functions using Eq.~\eqref{eq:delta}, integrals in Eq.~\eqref{eq:virint} can be reduced to corresponding simpler  MIs, the number of which is 66. DEs for these MIs can also be set up.

As for real corrections, asymptotic expansion of virtual-correction MIs at any point $z=z_0$ can be obtained in the form of Eq.~\eqref{eq:miexp} with the help of DEs, once we have boundary conditions for these DEs. The DEs have 6 singularities in the complex-$z$ plane, $z=0,\pm1,2,2(\pm\sqrt{2}-1)$, as shown in Fig.~\ref{fig:depole}. For the physical region, the relevant poles are $z=0, 2(\sqrt{2}-1), 1$, among which $z=2(\sqrt{2}-1)$ is a removable singularity. We will discuss in Appendix \ref{sec:singular} that this removable singularity does not affect the radius of convergence of asymptotic expansion at $z=1$, although it can decrease the precision if we estimate values for $z<2(\sqrt{2}-1)$ from the asymptotic expansion at $z=1$. The later problem has no impact if boundary conditions can be calculated to sufficient high precision, which is indeed the case as we will explain later.  Therefore, virtual-correction MIs in regions $0\sim1/4$, $1/4\sim3/4$ and $3/4\sim1$ can be respectively estimated by the asymptotic expansions of MIs at $z=0,1/2$ and $1$, where we introduce an expansion at a non-singular point $z=1/2$ so that the combination of real corrections and virtual corrections can be expressed by a single piecewise function.

Finally, let us discuss how to obtain boundary conditions for DEs of virtual-correction MIs. We find that, if we choose boundary conditions at $z\to1$, calculation of these  MIs either analytically or numerically to high precision is very hard.
The method proposed in Ref.\cite{Liu:2017jxz,Liu:2018dmc} provides a way to calculate MIs numerically to very high precision at any non-singular point $z$, which we will explain in Appendix \ref{sec:BCvirtual}. With this method, we can not only provide boundary conditions for DEs, but also do a self-consistent check. To this purpose, we use this method to calculate MIs at two points, say $z=z_1$ and $z_2$. With results at $z=z_1$ as boundary conditions, the DEs can give prediction for MIs at $z=z_2$, and the later values can be compared with the values obtained by this method. In our work, We have done this self-consistent check, and find a perfect agreement.

\section{Renormalization} \label{sec:renorm}
Bare quantities of fields $\Psi_b$ and $A_b$, coupling constant $g_{sb}$, and heavy quark mass $m_{Qb}$ are related to corresponding renormalized ones by the renormalization constants $\delta_2\,,\delta_3\,,\delta_g$ and $\delta_m$,
\begin{equation}
	\Psi_b = (1+\delta_2)^{1/2} \Psi \, , \
	A_b^{\mu} = (1+\delta_3)^{1/2} A^{\mu} \, , \
	g_{sb} =  (1+\delta_g) g_s \, , \
	m_{Qb} = (1+\delta_m) m_Q \,.
\end{equation}
In this paper, we choose $\msbar$ renormalization scheme for the coupling constant, and choose on-shell renormalizaiton scheme for gluon field, heavy quark field and heavy quark mass. It is convenient to rescale the renormalization constants as following
\begin{equation}
	\delta_i = \frac{\alpha_s}{\pi} \Gamma(1+\epsilon)
				\left( \frac{\pi \mu_r^2}{m_Q^2} \right)^{\epsilon}
				\hat{\delta} _i \, ,
\end{equation}
with
\begin{align}\label{eq:zren}
\begin{split}
	\hat{\delta}_2 & = -\frac{C_F}{4}
						\left( \frac{1}{\epsilon_{\uv}} +\frac{2}{\epsilon_{\ir}} + 4 + 6 \ln 2 \right) \,, \\
	\hat{\delta}_3 & = \left( \frac{5}{12} N_c - \frac{1}{6} n_f \right)
						\left( \frac{1}{\epsilon_{\uv}} -\frac{1}{\epsilon_{\ir}} \right) \,, \\
	\hat{\delta}_g & = -\frac{b_0}{4}
						\left( \frac{1}{\epsilon_{\uv}} - \ln \frac{\mu_r^2}{4m_Q^2} \right) \,, \\
	\hat{\delta}_m & = -\frac{3C_F}{4}
						\left( \frac{1}{\epsilon_{\uv}} + \frac{4}{3} + 2 \ln 2 \right) \,,											
\end{split}
\end{align}
where $b_0 = (11N_c-2n_f)/6$.

Summing over all counter terms, we obtain
\begin{equation}
	\int \ud \Phi_{\mathrm{Born}} \left( \delta_2 + 2\delta_g + \frac{\delta_m}{2 k\cdot P} \right) |\mathcal{M}_{\mathrm{LO}}|^2 \,,
\end{equation}
where $|\mathcal{M}_{\mathrm{LO}}|^2$ is the squared amplitude at LO in $\alpha_s$, and $\ud \Phi_{\mathrm{Born}}$ is defined in Eq.~\eqref{eq:phborn}.
This integral can be calculated easily.

Besides, we need to renormalize the operator defining the FF.
In $\msbar$ scheme, the counter term gives
\begin{equation}
	d_{\mathrm{Operator}}^{[1/8]} (z) =
		-\frac{\alpha_s}{2\pi}
		\frac{\Gamma(1+\epsilon)}{\epsilon}
		\left( \frac{4 \pi \mu_r^2}{\mu_f^2} \right)^{\epsilon}
		\int_z^1 \frac{\ud y}{y} P_{gg} (y)
			d_{\mathrm{LO}}^{[1/8]} \left(\frac{z}{y} \right) \, ,
\end{equation}
where $\mu_f$ is the factorization scale, $d_{\mathrm{LO}}^{[1/8]}(z)$ are given respectively in Eq.~\eqref{eq:lo1S01} and Eq.~\eqref{eq:lo1S08}, and the Altarelli-Parisi splitting function $P_{gg} (z)$ is
\begin{equation}
	P_{gg} (z) = b_0 \, \delta (1-z) +
				 2 N_c \left( \frac{z}{(1-z)_+} + \frac{1-z}{z} + z(1-z) \right) \, .
\end{equation}

\section{Results and discussion} \label{sec:summary}

\subsection{Final results} \label{sec:analytical}

Summing over real corrections, virtual corrections and all counter terms, we obtain finite results at NLO for both FFs. The results can be expressed in terms of piecewise functions,
\begin{align}
\begin{split}
	d_{\mathrm{NLO}}^{[1]} (z) & = \frac{\alpha_s^3}{2 \pi N_c m_Q^3 } \times
  		\left(d^{[1]}(z) 
  		+ \ln \left(\frac{\mu_r^2}{4m_Q^2} \right) b_0 \, d_{\mathrm{LO}}^{(0)} (z) 
  		+ \ln \left(\frac{\mu_f^2}{4m_Q^2} \right) f(z)
  		\right) \, , \\
  	d_{\mathrm{NLO}}^{[8]} (z) &= \frac{\alpha_s^3 (N_c^2-4)}{4\pi N_c(N_c^2-1) m_Q^3 } \times
  		\left(d^{[8]}(z)
  		+ \ln \left(\frac{\mu_r^2}{4m_Q^2} \right) b_0 \, d_{\mathrm{LO}}^{(0)} (z) 
  		+ \ln \left(\frac{\mu_f^2}{4m_Q^2} \right) f(z)
  		\right) \, ,
\end{split}
\end{align}
where $b_0$ is given below Eq.~\eqref{eq:zren}, $d_{\mathrm{LO}}^{(0)} (z)$ is given in Eq.~\eqref{eq:lo0},
\begin{align}
\begin{split}
	f(z) = & -\frac{n_f}{6} d_{\mathrm{LO}}^{(0)} (z)
			+ N_c \bigg(-2 (z+2) \text{Li}_2(z)
			-2 (z-1) \ln ^2(1-z)
			+2 (z-1) \ln(z) \ln (1-z)
			+(z-4) z \ln (z) \\
		&	-\frac{(2 z+1) \left(9 z^2-5 z-6\right) \ln (1-z)}{6 z}
			+\frac{46 z^3+\left(8 \pi ^2-3\right) z^2+4 \left(\pi ^2-9\right) z+4}{12 z}
			\bigg) \, ,
\end{split}
\end{align}
and
\begin{equation}\label{eq:res}
  	d^{[1/8]}(z) = 
  	\left\{
 	\begin{aligned}
 		&-\frac{N_c}{2 z} + \sum_{i=0}^{2} \sum_{j=0}^{\infty} \ln^i z \, (2z)^j \left(A_{ij}^{f} \, n_f + A_{ij}^{[1/8]} \, N_c +  \frac{A_{ij}^{N}}{N_c} \right) \,, & \text{for } 0<z<\frac{1}{4} \\
 		&\sum_{j=0}^{\infty} (2z-1)^j \left(B_{j}^{f} \, n_f + B_{j}^{[1/8]} \, N_c + \frac{B_{j}^{N}}{N_c} \right) \,, & \text{for }    \frac{1}{4} \le z \le \frac{3}{4} \\
 		&\sum_{i=0}^{3} \sum_{j=0}^{\infty} \ln^i (1-z) \, (2-2z)^j \, \left(C_{ij}^{f} \, n_f + C_{ij}^{[1/8]} \, N_c + \frac{C_{ij}^{N}}{N_c} \right) \,, & \text{for } \frac{3}{4}<z<1 
 	\end{aligned}
 	\right. \,.
\end{equation}
The coefficients $A_{ij}^{k},B_{j}^{k},C_{ij}^{k}$ can be evaluated numerically to very high precision, then analytical results can be obtained by fitting numerical results using PSLQ algorithm. For example, with $20$-digit precision, we get
 \begin{align}
\begin{split}
A_{00}^{[1]}=A_{00}^{[8]} &= 17-\frac{11 \zeta (3)}{8}-\frac{13 \pi ^2}{12}+\ln^2 2-\frac{\pi ^2}{4} \ln 2 \, .
\end{split}
\end{align} 
In practice, however, numerical results with high precision will be sufficient. In Appendix \ref{sec:coeff}, we present these coefficients up to $j=50$ with $18$ digits 
for each coefficient. With these numerical results, we can calculate $d_{\mathrm{NLO}}^{[1/8]} (z)$ to more than $15$-digit precision for any value of $z$.
To obtain about $150$-digit precision for any value of $z$, we will attach an ancillary file for the arXiv preprint in future, in which these coefficients will be calculated up to $j=500$ with $150$ digits for each coefficient.

\subsection{Numerical results}

\begin{figure}[htb!]
 \begin{center}
 \includegraphics[width=0.6\textwidth]{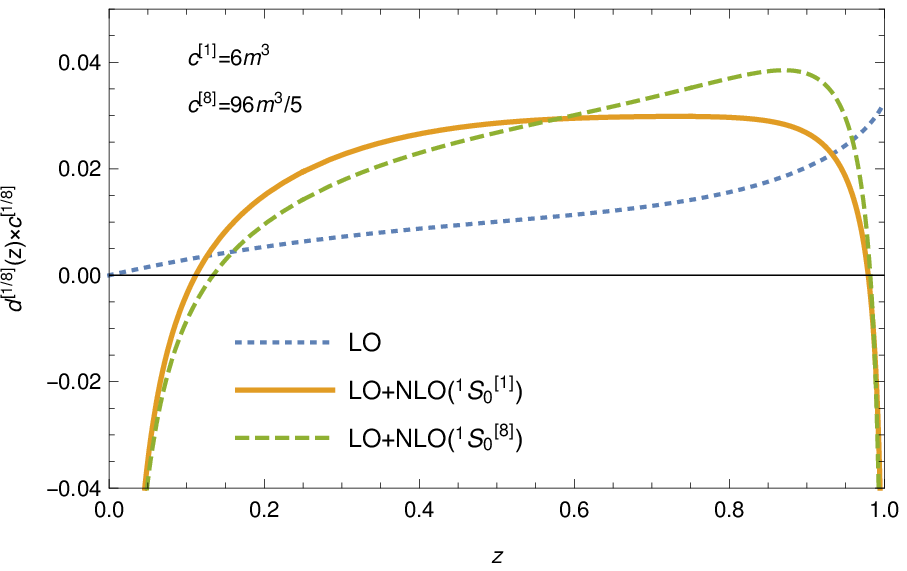}
  \caption{SDCs of the fragmentation functions of $g\to b\bar{b} (\CSaSz) $ and $g\to b\bar{b} (\COaSz) $ at LO and NLO. The dotted line is for $d^{[1]}_{\mathrm{LO}}(z)\times (6m_b^3)$ or $d^{[8]}_{\mathrm{LO}}(z)\times (96m_b^3/5)$, the solid line is for $(d^{[1]}_{\mathrm{LO}}(z)+ d^{[1]}_{\mathrm{NLO}}(z)) \times (6m_b^3)$ and the dashed line is for $(d^{[8]}_{\mathrm{LO}}(z)+ d^{[8]}_{\mathrm{NLO}}(z)) \times (96m_b^3/5)$, with scale choices $\mu_r=\mu_f=2m_b$. The superscript $[1]$ or $[8]$ respectively denotes the color-singlet or color-octet states of $b\bar{b}$. \label{fig:restotal}}
 \end{center}
\end{figure}

To see the effects of NLO corrections, we choose parameters the same as that in Ref.\cite{Artoisenet:2014lpa}, with $m_b=4.75\gev$, $N_c=3$, $n_f=4$, and $\alpha_s (\mu_r=2m_b)=0.181 $. In Fig.~\ref{fig:restotal},
we plot the curves of LO FFs and LO+NLO FFs with $\mu_r = \mu_f = 2m_b$. To show color-singlet FFs and color-octet FFs in the same figure, we introduce overall factors $c^{[1]}=6m^3$ and $c^{[8]}=96m^3/5$ for them, respectively. We find that our result of NLO FF of $g\to Q\bar{Q} (\CSaSz) +X$ has some differences from that obtained in Ref.\cite{Artoisenet:2014lpa}, especially when $z\rightarrow 0$. With $\mu_r = \mu_f = 2m_b$, we also provide K-factors (the ratio of LO+NLO over LO) of some special values of $z$ in Tab. \ref{table:Kfactor}, where we find that K-factors are very significant for most values of $z$.

\begin{table}[htb!]
	\begin{tabular*}{\textwidth}{@{\extracolsep{\fill}} ccc|ccc}
		\hline \hline
$z$ & $K^{[1]}$ & $K^{[8]}$ & $z$ & $K^{[1]}$ & $K^{[8]}$ \\
\hline
$0.05$ & $-22.2154523436534$ & $-24.6733986813826$ & $0.55$ & $2.72527357573690$ & $2.66250417113448$ \\
$0.10$ & $-1.19896707966308$ & $-2.87199122364689$ & $0.60$ & $2.59460446402429$ & $2.64623824982464$ \\
$0.15$ & $1.96212951637830$ & $0.686093914799466$ & $0.65$ & $2.44998117223888$ & $2.61539737442995$ \\
$0.20$ & $2.80788837290754$ & $1.79077857153724$ & $0.70$ & $2.28930766059255$ & $2.56549304190537$ \\
$0.25$ & $3.06753043346018$ & $2.24294113774066$ & $0.75$ & $2.10880071058012$ & $2.48839147107813$ \\
$0.30$ & $3.12597786090724$ & $2.45762890746927$ & $0.80$ & $1.90118848019761$ & $2.36993271046623$ \\
$0.35$ & $3.10100074972157$ & $2.56808157787236$ & $0.85$ & $1.65072948967985$ & $2.18335185168051$ \\
$0.40$ & $3.03565411189113$ & $2.62635170085241$ & $0.90$ & $1.31711318345314$ & $1.86698847168032$ \\
$0.45$ & $2.94726594486785$ & $2.65539830292573$ & $0.95$ & $0.755737935988107$ & $1.20834068587147$ \\
$0.50$ & $2.84294512935356$ & $2.66590375106102$ & $0.99$ & $-0.694039121672193$ & $-0.839542885587686$ \\
\hline \hline
	\end{tabular*}
	\caption{K-factors at different values of $z$. The superscript $[1]$ or $[8]$ respectively denotes the color-singlet or color-octet states of $b\bar{b}$. \label{table:Kfactor}}
\end{table}

As shown in Eq.~\eqref{eq:res}, NLO FFs are negative and divergent at both $z=0$ and $z=1$, with leading divergence $1/z$ at $z=0$. Thus total fragmenting probabilities obtained by integrating NLO FFs over $z$ from 0 to 1 are infinite.
As cross sections are obtained by convoluting FFs with smooth functions of $z$, they only sensitive to a little higher moments of FFs,
\begin{equation}
	\int_0^1 \ud z \, z^n (d^{[1/8]}_{\mathrm{LO}}(z)+ d^{[1/8]}_{\mathrm{NLO}}(z))*c^{[1/8]} \, ,
\end{equation}
numerical results of which are shown in Table.~\ref{table:probability} for $n=2,4,6$. We find that, unlike K-factors for fixed $z$, K-factors of 4-th moments and 6-th moments are moderate.
\begin{table}[htb!]
	\begin{tabular*}{\textwidth}{@{\extracolsep{\fill}} c c c c c }
		\hline \hline
		state          			 & SDCs$*c^{[1/8]}$ 					& $z^2$ & $z^4$ & $z^6$ \\
		\hline
		& LO ($\times 10^{-3}$)                               & $5.55116944444444$ & $3.85331761904762$ & $2.99519856859410$ \\
		\hline
		\multirow{2}{*}{$\CSaSz$} & LO+NLO ($\times 10^{-3}$) & $7.54577896198438$ & $3.90413390635734$ & $2.31890675629641$ \\
		& K-factor  					                      & $1.35931339107945$ & $1.01318767159461$ & $0.774208021001048$ \\
		\hline
		\multirow{2}{*}{$\COaSz$} & LO+NLO ($\times 10^{-3}$) & $8.94021475091022$ & $4.99398540595690$ & $3.12511443982943$ \\
		& K-factor  					                      & $1.61051015292958$ & $1.29602225917499$ & $1.04337471064441$ \\
		\hline \hline
	\end{tabular*}
	\caption{Moments and K-factor of SDCs.  \label{table:probability}}
\end{table}

\begin{figure}[htb!]
 \begin{center}
 \includegraphics[width=0.6\textwidth]{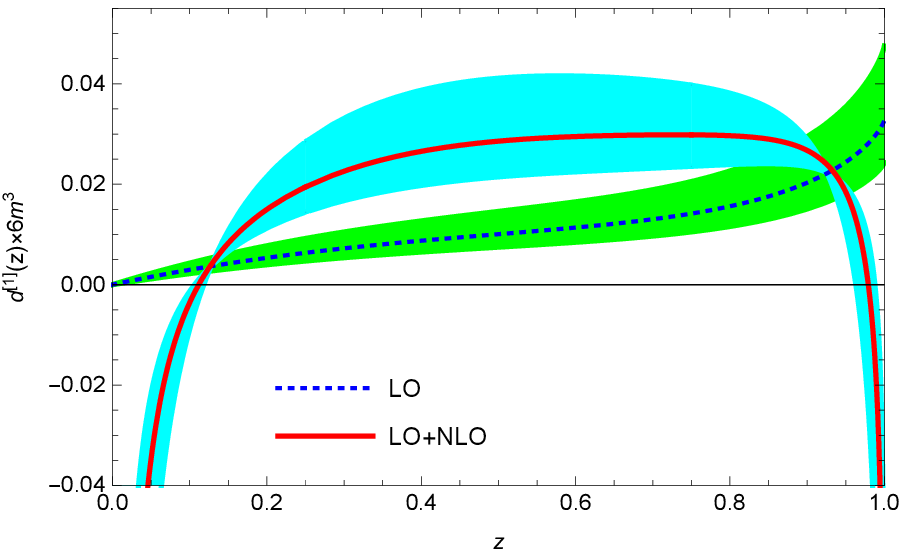}
  \caption{SDCs of the fragmentation function of $g\to b\bar{b} (\CSaSz) +X$ at LO and NLO. The dotted line is for $d^{[1]}_{\mathrm{LO}}(z)\times (6m_b^3)$, while the solid line is for $(d^{[1]}_{\mathrm{LO}}(z)+ d^{[1]}_{\mathrm{NLO}}(z)) \times (6m_b^3)$, with scale choices $\mu_r=\mu_f=2m_b$. The bands are obtained by varying the renormalization scale $\mu_r$ by a factor of 2.  \label{fig:res1S01ren}}
 \end{center}
\end{figure}

\begin{figure}[htb!]
 \begin{center}
 \includegraphics[width=0.6\textwidth]{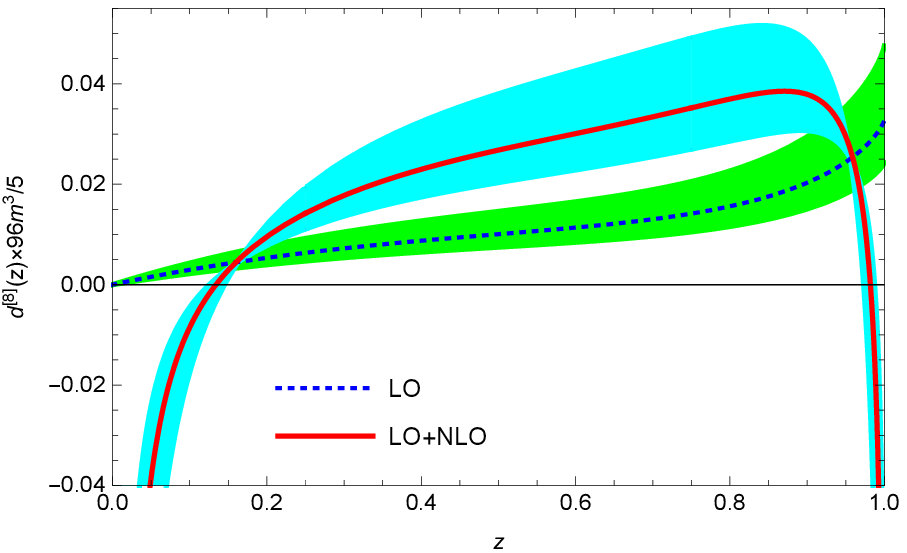}
  \caption{SDCs of the fragmentation function of $g\to b\bar{b} (\COaSz) +X$ at LO and NLO. The dotted line is for $d^{[8]}_{\mathrm{LO}}(z)\times (96m_b^3/5)$, while the solid line is for $(d^{[8]}_{\mathrm{LO}}(z)+ d^{[8]}_{\mathrm{NLO}}(z)) \times (96m_b^3/5)$, with scale choices $\mu_r=\mu_f=2m_b$. The bands are obtained by varying the renormalization scale $\mu_r$ by a factor of 2. \label{fig:res1S08ren}}
 \end{center}
\end{figure}

The sensitivity of LO and LO+NLO FFs with respective to the renormalization scale $\mu_r$ is illustrated in Fig.~\ref{fig:res1S01ren} for $g\to Q\bar{Q} (\CSaSz) +X$ and Fig.~\ref{fig:res1S08ren} for $g\to Q\bar{Q} (\COaSz) +X$, with $\mu_f=2m_b$ and varying $\mu_r$ from $m_b$ to $4m_b$.
We find that theoretical uncertainties are still large with NLO corrections.

\begin{acknowledgments}

We thank Hao-Yu Liu and Yu-Jie Zhang for useful discussions, and thank Feng Feng and Yu Jia for useful communications.
The work is supported in part by the National Natural Science Foundation
of China (Grants No. 11475005 and No. 11075002), the National Key Basic Research Program
of China (No. 2015CB856700), and High-performance Computing Platform of Peking University.

\end{acknowledgments}

\vspace{1.0 cm}

$Note~ added.$ --- While this paper was being finalized, two related preprints appeared \cite{Artoisenet:2018dbs,Feng:2018ulg}. In Ref.~\cite{Artoisenet:2018dbs}, the authors calculated NLO corrections for FF of $g\to Q\bar{Q}(\state{1}{S}{0}{8})+X$ using FKS subtraction method; while in Ref.~\cite{Feng:2018ulg}, the authors calculated NLO corrections for FFs of $g\to Q\bar{Q}(\state{1}{S}{0}{1})+X$ and $g\to Q\bar{Q}(\state{1}{S}{0}{8})+X$ using sector decomposition. Our high-precision results agree with K-factors obtained in these two works within their estimated errors.


\appendix

\section{IBP reduction with unregularized rapidity divergence}
\label{sec:reg}

If all integrals are well regularized by dimensional regularization, IBP reduction relations should be independent of the infinitesimal imaginary parts $i \eta$, which means that coefficients of the relations are the same no matter a denominator is $E_j + i \eta$ or $E_j -  i\eta$. This is the reason why we ignore the infinitesimal imaginary parts when using IBP reduction.

However, in this paper we encounter some integrals that can not be regularized by dimensional regularization only.
There is a MI in the calculation of real correction
\begin{equation}\label{eq:regequ}
\int \ud \Phi_{\mathrm{real}} \frac{1}{E_1 \, E_4} \, ,
\end{equation}
which equals to
\begin{equation}
\frac{1}{(4 \pi)^2 z^2} \int_{0}^{1} \frac{\mathrm{d} z_{1}}{z_{1}}
\int \frac{\mathrm{d}^{D-2}  k_{1\perp}}{(2\pi)^{D-2}} \frac{\mathrm{d}^{D-2} k_{2\perp}}{(2\pi)^{D-2}} 
\frac{1}{\left(k_{2\perp}- k_{1\perp}\right)^2 \left(k_{1\perp}^2 + \left(\frac{1-z}{z}\right)^2 z_1(1-z_1) + \frac{1-z}{z} (1-z_1) \right)} \, ,
\end{equation}
where we integrated out $k_1^-,k_2^-$ and $k_2^+$, denoted $k_1^+ = (1-z)z_1 P_c^+$, and did the replacement
\begin{equation}
k_{1\perp} \rightarrow \sqrt{\frac{z_1}{1-z_1}} \, k_{1\perp} \, , \
k_{2\perp} \rightarrow \sqrt{\frac{1-z_1}{z_1}} \, k_{2\perp} \, .
\end{equation}
It is clear now that the integration over $z_1$ is divergent at $z_1=0$ and it can not be regularized by dimensional regularization. This divergence is usually called rapidity divergence, and it is in fact well-known that it cannot be regularized by dimensional regularization. Similar problem exists when changing $E_4$ to $E_5$. Because the MI in Eq.~\eqref{eq:regequ} is unregularized, on  one hand we do not know how to calculate it, and on the other hand the IBP reduction which expresses SDCs as linear combination of MIs may give  wrong result. 

To explain why IBP reduction can be wrong, let us replace $E_1,E_4$ in Eq.~\eqref{eq:edef} respectively by
\begin{equation}
E_1' = (k_1 + k_2)^2 \, , \
E_4' = (k_1 + P)^2 \, , \
\end{equation}
which does not change the integral because of $\delta$ functions in the definition of $\ud \Phi_{\mathrm{real}}$. If we then simply replace these $\delta$ functions by propagator denominators using Eq.~\eqref{eq:delta} and perform IBP reduction of original expression by ignoring the infinitesimal imaginary parts, we find two equal loop integrals,
\begin{equation} \label{eq:wmi1}
\frac{P \cdot n}{z^{2} 2!}
\int \frac{\mathrm{d}^{D} k_1}{(2\pi)^{D}}
\frac{\mathrm{d}^{D} k_2}{(2\pi)^{D}}
\frac{1}{E_1' E_4' E_{12} E_{13} E_{14} } \, ,
\end{equation}
and
\begin{equation} \label{eq:wmi2}
\frac{1-2\epsilon}{\epsilon} 
\frac{(P \cdot n)^2}{z^{2} 2!}
\int \frac{\mathrm{d}^{D} k_1}{(2\pi)^{D}}
\frac{\mathrm{d}^{D} k_2}{(2\pi)^{D}}
\frac{1}{E_1' E_4' E_8 E_{13} E_{14} } \, .
\end{equation}
Because they are equal to each other, we can choose either the former or the later as our MI.
However, on the other hand, once we replace propagator denominators back to $\delta$ functions, the Eq.~\eqref{eq:wmi2} will vanish as it lacks of $E_{12} $, while the Eq.~\eqref{eq:wmi1} will be changed to MI in Eq.~\eqref{eq:regequ}. Therefore, the final results are ambiguous.

To get unambiguous results both in the reduction step and in the calculation of MIs, we in principle need all involved integrals to be well regularized. 
We thus introduce an additional regulator besides spacetime dimension $D=4-2\epsilon$, and take the limit of this new regulator to zero before take the limit of $\epsilon\to0$. In this way, divergences that are regularized by dimensional regularization will not be affected. A possible choice of the new regulator is gluon mass in the phase space integration, which means we use
\begin{equation}
\ud \Phi'  =
\frac{P \cdot n}{z^{2} 2!}
\frac{\mathrm{d}^{D} k_1}{(2\pi)^{D-1}}
\frac{\mathrm{d}^{D} k_2}{(2\pi)^{D-1}}
\delta _+ (k_1^{2}-m_g^2)
\delta _+ (k_2^{2}-m_g^2)
\delta \left( k_1 \cdot n + k_2 \cdot n - \frac{1-z}{z} P \cdot n \right) \, ,
\end{equation}
instead of $\ud \Phi$. Note that, although gluon mass should also be introduced in Feynman amplitudes to be self-consistent, it is easy to show that only the gluon masses in phase space integration have non-vanishing effect. With this regulator, we find all involved integrals in our calculation are well regularized, and thus the IBP reduction do not introduce any ambiguity. After the IBP reduction and then take the limit $m_g\to0$ in any place as far as the operation does not result in unregularized integrals, $m_g$ still presents in four MIs 
\begin{equation}\label{eq:mieta}
\int \ud \Phi' \frac{1}{E_1 \, E_4} \, ,
\int \ud \Phi' \frac{1}{E_1 \, E_4^2} \, ,
\int \ud \Phi' \frac{P\cdot n}{E_1 \, E_4 \, E_9} \, ,
\int \ud \Phi' \frac{P\cdot n}{E_1 \, E_4 \, E_{10}} \, ,
\end{equation}
besides the other four MIs obtained by changing $E_4$ to $E_5$.

As an example, we calculate the first MI in Eq.\eqref{eq:mieta} in the limit of $m_g \rightarrow 0$.
After we integrate out $k_1^-$, $k_2^-$, $k_2^+$, $k_{2\perp}$ and $k_{1\perp}$ sequentially, we get
\begin{equation}
m_g^{-2\epsilon} \Gamma(\epsilon)^2 z^{-2}
\int_0^1 \ud z_1 \, z_1^{-1+\epsilon}
(1-2z_1+2z_1^2)^{-\epsilon}
(t^2 z_1 + t + m_g^2 / z_1)^{-\epsilon} \, ,
\end{equation}
where $t=(1-z)/z$. Because of dimensional regularization, only the region $z_1 \sim m_g^2$ survives in the limit of $m_g \rightarrow 0$.
So we set $z_1=m_g^2 \, y$ and take limit of $m_g \rightarrow 0$, then we get
\begin{equation} \label{eq:regequval}
\Gamma(\epsilon)^2 z^{-2}
\int_0^{\infty} \ud y \, y^{-1+\epsilon}
(t + 1 / y)^{-\epsilon}  =
	z^{-2+2\epsilon} (1-z)^{-2\epsilon} \Gamma(2\epsilon) \Gamma(\epsilon) \Gamma(-\epsilon)\, .
\end{equation}
It tells us that, after introducing and then removing the gluon mass regulator, the MI in Eq.~\eqref{eq:regequ} is eventually well regularized by dimensional regularization. We can similarly calculate the other three MIs in Eq.~\eqref{eq:mieta}, and find that they can be obtained from the first MI by multiplying factors $2\epsilon\,,1$ and $z/(1-z) $, respectively.

Before describing how to apply the above method to our problem, let us first examine the following integral
\begin{equation} \label{eq:test}
\int \ud \Phi \frac{1}{E_1 E_4 E_7} \, ,
\end{equation}
which is well regularized by dimensional regularization~\footnote{It is well-regularized only if we first integrate out transverse momentum before integrate "+" momentum.}, and thus we can calculate it numerically without introducing any other regulator. On the other hand, again without introducing any other regulator, we use IBP naively to reduce it to MIs. We find the reduced result is unique. All MIs obtained here except the one in Eq.~\eqref{eq:regequ} are well regularized by dimensional regularization, which can be easily evaluated. Then if we replace the unregularized MI by Eq.~\eqref{eq:regequval}, we find the numerical result of Eq.~\eqref{eq:test} agrees with the value calculated by applying IBP reduction. This test tells us two things. The first is that our gluon mass regulator can indeed give correct result. We thus take Eq.~\eqref{eq:regequval} as the value of MIs defined in Eq.~\eqref{eq:regequ}, and similarly for other unregularized MIs. The second is that, if the original express is well regularized by dimensional regularization, using IBP naively may have no problem.

Based on the above lessons, we divide our original express two parts. The first part is well regularized by dimensional regularization, which is then reduced to MIs by using IBP naively. We check this part numerically and find good agreement between results before and after the IBP reduction. As the second part is unregularized, we introduce the above gluon mass regulator before applying IBP. After inserting the values of MIs, we find the second part in our decomposition eventually vanishes.

\section{Removable singularites and their effects} 
\label{sec:singular}

In this work, we encounter some removable singularities. Some of them determine the convergence radius of asymptotic expansion at some points, and others only decrease the precision of higher order coefficients in the asymptotic expansion. In the following discussion, to be definite we discuss the case where there is a removable singularity at $z=1/2$ and there is a non-removable singularity at $z=1$. We will do asymptotic expansion at $z=0$.

Let us first discuss the case where there are more than one analytical structure at $z=0$, and the singularity at $z=1/2$ is removable only after the summation of contributions from all structures. Here is an example,
\begin{equation}
f(z)=\frac{\frac{\ln z}{1-z} + 2 \ln 2}{1-2z},
\end{equation}
where $z=1/2$ is indeed a removable singularity. When we do the asymptotic expansion at $z=0$, we get two series, where one comes from the analytical part and the other one comes from the part proportional to $\ln z$. As $z=1/2$ is a non-removable singularity of each of the two parts, the convergence radius of each of the series is $1/2$. Although $z=1/2$ becomes a removable singularity in the summation of two parts, the convergence radius of the asymptotic expansion at $z=0$ is still $1/2$. The reason is that we have no way to reorganize the two series to a single series so that it is convergent everywhere in $1/2<|z|<1$.
In our calculation, MIs in real corrections have this kind of removable singularity at $z=1/2$, which determine convergence radius of asymptotic expansion at both $z=0$ and $z=1$.

Now let us discuss the case where  $z=1/2$ is a removable singularity for each non-vanishing analytical structure at $z=0$. A special case is that there is only one non-vanishing analytical structure, for example
\begin{equation}
g(z)=\frac{2\ln (2-2z)}{1-2z} \, ,
\end{equation}
where $1/2$ is a removable singularity while $1$ is a branch point. If we denote 
\begin{equation}
g(z)= \sum_{n=0}^{\infty} a_n z^n \,,
\end{equation}
we have
\begin{equation} \label{eq:anser}
a_n= 2^{n+1} \ln2 - \sum_{i=0}^{n-1} \frac{2^{i+1}}{n-i} \,,
\end{equation}
based on which we can calculate the convergence radius: $\lim_{n\to\infty} \frac{a_n}{a_{n+1}}=1$. We thus find that the singularity $z=1/2$ does not affect the convergence radius at $z=0$. However, we will show that this singularity has other effects. To this purpose, we note that  $g(z)$ satisfies the following DE,
\begin{equation}
\left(\frac{1}{2}-z\right) \frac{\ud g(z)}{\ud z} = g(z) - \frac{1}{1-z} \,,
\end{equation}
with initial condition $g(0)=a_0=2\ln 2$.
The DE can generate the recursion relation
\begin{equation}
a_{n+1} = 2a_n - \frac{2}{n+1} \,,
\end{equation}
which determines higher order coefficients in the expansion. However, when we solve DEs numerically, the initial condition can have only finite precision. If we denote the absolute error of $a_0$ as $\lambda$, the absolute error of $a_n$ calculated from the recursion relation is $2^n\lambda$. At the point $z=x$, the contributed error from $a_n$ is $(2x)^n\lambda$, which is much larger than $\lambda$ if $x>1/2$ and $n$ is large. If we reduce this error by truncating the expansion to small $n$, then there will be a large systematic error at the order of $x^n$. The best accuracy at $z=x$ that one can obtain is to choose a truncation $n$ so that $(2x)^n\lambda\sim x^n$, which gives $n\sim\log_2 \lambda^{-1}$ and $x^n\sim \lambda^{-\log_2 x}$. For example, for the point $x=\sqrt{2}/2$, the smallest absolute error that we can get is $\lambda^{1/2}$, which is larger than the absolute error $\lambda$ at $x=0$.

In virtual corrections, if we do asymptotic expansion at $z=1$, the removable singularity at $z=2(\sqrt{2}-1) $ belongs to the second type, and the convergence radius is determined by the singularity at $z=0$. Let us denote $1-2(\sqrt{2}-1)=a^{-1}$, then the best accuracy at $z=1-x$ that we can obtained is determined by  $(a x)^n\lambda\sim x^n$, which gives $n\sim\log_a \lambda^{-1}$ and $x^n\sim \lambda^{-\log_a x}$. In this work, we want to estimate the value at $z=3/4$ from the expansion at $z=1$, which gives the best accuracy about $\lambda^{0.786}$. If we need the accuracy to be about $10^{-15}$, we find $\lambda\sim10^{-19}$ and $n\sim 25$, which means that we need initial condition for the expansion at $z=1$ to have four more significant digits.

\section{Boundary conditions of MIs in real corrections}
\label{sec:BCreal}

MIs in real corrections have the form
\begin{equation} \label{eq:realbc}
	\int \ud \Phi_{\mathrm{real}} \frac{1}{E_a^{n_a} \, E_b^{n_b} \, E_c^{n_c}\, E_d^{n_d}  } \, ,
\end{equation}
where $d\in\{8,9,10,11\}$, $a,b,c \in \{1,\dots,7\}$ and $\ud \Phi_{\mathrm{real}}$ is defined in Eq.~\eqref{eq:realint}. We calculate most MIs in the limit $z\to1$ in this Appendix, with the other MIs which are not regularized by dimensional regularization calculated in Appendix \ref{sec:reg}.
From Eq.~\eqref{eq:realint}, $\delta \left( k_1 \cdot n + k_2 \cdot n - \frac{1-z}{z} P \cdot n \right)$ together with conditions $k_1^+>0$ and $k_2^+>0$ requires that $k_1^+$ and $k_2^+$ must be at the order of $1-z$ when $z\rightarrow1$. Otherwise, if taking $k_1^+\ll (1-z)P^+ $ as an example, the integral will be proportional to 
\begin{equation}
	\int_0^{\infty} \ud k_1^+ \, (k_1^+)^{a+b\epsilon} \, ,
\end{equation}
which equals $0$ in dimensional regularization.
Introducing the parametrization  $k_1^+ = (1-z)z_1 P_c^+$ and $k_2^+ = (1-z)(1-z_1) P_c^+$, $\ud \Phi_{\mathrm{real}}$ becomes
\begin{equation}
	\int \ud \Phi_{\mathrm{real}} =
	\frac{1}{(4 \pi)^2 z (1 -z ) 2!} \int_{0}^{1} \frac{\mathrm{d} z_{1}}{z_{1} (1 - z_{1})}
    		\int \frac{\mathrm{d}^{D-2} k_{1\perp}}{(2\pi)^{D-2}} \frac{\mathrm{d}^{D-2} k_{2\perp}}{(2\pi)^{D-2}} \, .
\end{equation}
In the limit of $z\rightarrow1$, $E_i$ given in Eq.~\eqref{eq:edef} become
\begin{align} \label{eq:elim}
\begin{split}
	\hat{E}_1 	& = \frac{1}{2} \left(\frac{1-z_1}{z_1} k_{1\perp}^2 + \frac{z_1}{1-z_1} k_{2\perp}^2 - 2 k_{1\perp} \cdot k_{2\perp} \right) \, , \\
	\hat{E}_2 	& = \frac{1}{2\lambda} \left(\frac{k_{1\perp}^2}{z_1} + \lambda^2 z_1 \right) \, , \\
	\hat{E}_3 	& = \frac{1}{2\lambda} \left(\frac{k_{2\perp}^2}{1-z_1} + \lambda^2 (1-z_1) \right) \, , \\
	\hat{E}_4 	& = \frac{1}{\lambda} \left(\frac{k_{1\perp}^2}{z_1} + \lambda \right) \, , \\
	\hat{E}_5 	& = \frac{1}{\lambda} \left(\frac{k_{2\perp}^2}{1-z_1} + \lambda \right) \, , \\
	\hat{E}_6 	& = \frac{1}{2\lambda} \left(\frac{k_{1\perp}^2}{z_1} + \frac{k_{2\perp}^2}{1-z_1} + \lambda^2 \right) \, , \\
	\hat{E}_7 	& = \frac{1}{\lambda} \left(\frac{k_{1\perp}^2}{z_1} + \frac{k_{2\perp}^2}{1-z_1} + \lambda \right) \, ,
\end{split}	
\end{align}
where $\lambda=1-z$.
As $\lambda\to0$, each MI has at most four nonvanishing regions,
\begin{align}
\begin{split}
	&k_{1\perp}^2 \sim \lambda^2 \, , k_{2\perp}^2 \sim \lambda^2 \, ; \\
	&k_{1\perp}^2 \sim \lambda^2 \, , k_{2\perp}^2 \sim \lambda \, ; \\
	&k_{1\perp}^2 \sim \lambda \, , k_{2\perp}^2 \sim \lambda^2 \, ; \\
	&k_{1\perp}^2 \sim \lambda \, , k_{2\perp}^2 \sim \lambda \, . \\
\end{split}
\end{align}
To obtain boundary conditions, we only need to calculate the leading contribution in each region, which is proportional to $\lambda^{n \epsilon}$ with $n=-2,-3$ or $-4 $.
The calculation is a little different depending on whether $E_1$ presents in Eq.~\eqref{eq:realbc}.

Without $E_1$, there is no cross term $k_{1\perp} \cdot k_{2\perp}$ in the limit $\lambda\to0$. We thus first rescale momenta by
\begin{equation}
	k_{1\perp} \rightarrow \sqrt{z_1} \, k_{1\perp} \, , \
	k_{2\perp} \rightarrow \sqrt{1-z_1} \, k_{2\perp} \, ,
\end{equation}
and then integrate out $k_{2\perp}$.
After that, the integration over $k_{1\perp}$ is very simple unless it has the form
\begin{equation}
	\int \frac{\mathrm{d}^{D-2} k_{1\perp}}{(2\pi)^{D-2}} \frac{1}{(k_{1\perp}^2+1)^{n_1}(k_{1\perp}^2+a(z_1))^{n_2}} \, ,
\end{equation}
with $a(z_1) \neq 0,1$.
For this kind of integrals, we integrated out $k_{1\perp}$ after Feynman parametrization.
Finally, the integration over $z_1$ and Feynman parameters can be calculated analytically with the help of sector decomposition Ref.\cite{Binoth:2000ps,Heinrich:2008si}, which can isolate mixed divergences from parameter integrals.
There are two widely used programs that can do the sector decomposition, SecDec \cite{Carter:2010hi,Borowka:2012ii,Borowka:2012yc,Borowka:2013cma} and FIESTA \cite{Smirnov:2008py,Smirnov:2009pb,Smirnov:2013eza,Smirnov:2015mct}. We use SecDec in this paper.

If $E_1$ presents, we first rescale momenta by
\begin{equation}
	k_{1\perp} \rightarrow \sqrt{\frac{z_1}{1-z_1}} \, k_{1\perp} \, , \
	k_{2\perp} \rightarrow \sqrt{\frac{1-z_1}{z_1}} \, k_{2\perp} \, ,
\end{equation}
and then do the the replacement $k_{1\perp} \rightarrow k_{1\perp} + k_{2\perp}$, which changes $\hat{E}_1$ to $k_{1\perp}^2$ and moves the cross term $k_{1\perp} \cdot k_{2\perp}$ to other denominators.
To proceed, we introduce Feynman parametrization and integrate out $k_{2\perp}$. Then the integration of  $k_{1\perp}$ has the form
\begin{equation}
	\int \frac{\mathrm{d}^{D-2} k_{1\perp}}{(2\pi)^{D-2}} \frac{1}{(k_{1\perp}^2)^{n_1}(k_{1\perp}^2+a(z_1))^{n_2}} \, ,
\end{equation}
which can be easily integrated out.
Finally, integration of Feynman parameters can be worked out with the help of sector decomposition.

All analytical results of MIs calculated here have been checked by numerical results computed by SecDec, and good agreement is found.

\section{Calculation of MIs in virtual corrections}
\label{sec:BCvirtual}

MIs for virtual corrections in Eq.~\eqref{eq:virint} can be expressed as
\begin{equation} \label{eq:virbc}
\frac{1}{4 \pi z (1 -z )}
\int \frac{\mathrm{d}^{D-2} k_{\perp}}{(2\pi)^{D-2}}
\frac{\mathrm{d}^{D} l}{(2\pi)^{D}}
\prod_{i} \frac{1}{F_i^{\nu_i}} \, ,
\end{equation}
where $F_i$ are defined in Eq.~\eqref{eq:edefloop} with $k^2=0$, $k^-=k^2_{\perp}/(2k^+) $ and $k^+=(1-z)P^+/z $. We apply the method proposed in Ref.\cite{Liu:2017jxz,Liu:2018dmc} to calculate these MIs at any regular point $z=z_0$. To this purpose, we change  $F_i$ to $F_i+i \eta$ for $i\neq 1,2$ to obtain new MIs. We can set up DEs of the new MIs by first differentiating them with respect to $\eta$ and then reducing the obtained expressions to the new MIs using IBP reduction. If we also know boundary conditions of the new MIs at a special value of $\eta$, we can solve the DEs numerically to obtain the new MIs at $\eta=0^+$  with very high precision, which are nothing but our desired old MIs. 

The boundary that we choose is at $\eta \rightarrow \infty$. To calculate the boundary conditions, we first perform Feynman parameterization and then shift $l$ to remove cross terms. The obtained results are proportional to
\begin{equation} \label{eq:virmi}
		\iint \ud x_1 \dots \ud x_n \int \frac{\mathrm{d}^{D-2} k_{\perp}}{(2\pi)^{D-2}}
        \frac{\mathrm{d}^{D} l}{(2\pi)^{D}}
    	\frac{1}{(k_{\perp}^2+a)^{n_1} (l^2-b \, k_{\perp}^2-c+i\eta)^{n_2} \, (l \cdot n+d+i\eta)^{n_3} } \, ,
\end{equation}
where $b,c,d$ are functions of $z$ and the Feynman parameters $x_1,\dots,x_n$, and $a$ is a function of $z$.
As $\eta \rightarrow \infty$, there are only two regions for this integral,
\begin{align}
\begin{split}
&	l^2 \sim \eta \, , \ k_{\perp}^2 \sim 1 \,; \\
&	l^2 \sim \eta \, , \ k_{\perp}^2 \sim \eta \,.
\end{split}
\end{align}
The leading term of the first region gives
\begin{equation}
	\eta^{2-n_2-n_3-\epsilon} \, i^{-n_3} \iint \ud x_1 \dots \ud x_n \int \frac{\mathrm{d}^{D-2} k_{\perp}}{(2\pi)^{D-2}}
		\frac{1}{(k_{\perp}^2+a)^{n_1}}
	\int \frac{\mathrm{d}^{D} l}{(2\pi)^{D}}
		\frac{1}{(l^2+i)^{n_2} } \, ,
\end{equation}
which can be easily integrated out. The leading term of the second region gives
\begin{equation} \label{eq:virreg2}
	\eta^{3-n_1-n_2-n_3-2\epsilon} \, i^{-n_3} \iint \ud x_1 \dots \ud x_n \int \frac{\mathrm{d}^{D-2} k_{\perp}}{(2\pi)^{D-2}}
		\frac{1}{(k_{\perp}^2)^{n_1}}
	\int \frac{\mathrm{d}^{D} l}{(2\pi)^{D}}
		\frac{1}{(l^2-b \, k_{\perp}^2 +i)^{n_2} } \, ,
\end{equation}
the integrand of which is proportional to $b^{n_1-1+\epsilon}$ after integrating out $k_{\perp}$ and $l$.
Though $b$ is a function of Feynman parameters and $z$, the dependence of $z$ can be factorized out.
So the integration over Feynman parameters can be easily performed.

\section{Coefficients} \label{sec:coeff}
In this Appendix, we give the coefficients defined in Eq.\eqref{eq:res}. 
The coefficients of asymptotic expansion at $z=0$ with different powers of $\ln(z)$ are shown respectively in Table.\ref{table:coeffA2} $\sim$ \ref{table:coeffA0}.
The coefficients of asymptotic expansion at $z=1/2$ are shown in Table.\ref{table:coeffB}.
The coefficients of asymptotic expansion at $z=1$ with different powers of $\ln(1-z)$ are shown respectively in Table.\ref{table:coeffC3} $\sim$ \ref{table:coeffC0}.
To obtain $150$-digit precision for SDCs at any value of $z$, we will attach an ancillary file for the arXiv preprint in future, in which these coefficients will be calculated up to $j=500$ with $150$ digits for each coefficient.


\begin{table}[htb!]
\begin{tabular*}{\textwidth}{@{\extracolsep{\fill}}*{1}{*{5}{l}}}
\hline \hline
 $j$ & $2^j \, A_{2j}^f$ & $2^j \, A_{2j}^N$ & $A_{2j}^{[1]}$ & $A_{2j}^{[8]}$ \\ 
$0$ & $0$ & $0$ & $0$ & $0$ \\
$1$ & $0$ & $0$ & $-0.125000000000000000$ & $-0.500000000000000000$ \\
$2$ & $0$ & $0.0625000000000000000$ & $0.171875000000000000$ & $0.187500000000000000$ \\
$3$ & $0$ & $0$ & $0.406250000000000000$ & $0.437500000000000000$ \\
$4$ & $0$ & $0.0104166666666666667$ & $0.419921875000000000$ & $0.583333333333333333$ \\
$5$ & $0$ & $0.0125000000000000000$ & $0.613020833333333333$ & $0.830729166666666667$ \\
$6$ & $0$ & $0.0125000000000000000$ & $0.773209635416666667$ & $1.08658854166666667$ \\
$7$ & $0$ & $0.0119047619047619048$ & $0.964741443452380952$ & $1.38359375000000000$ \\
$8$ & $0$ & $0.0111607142857142857$ & $1.16243751162574405$ & $1.70711495535714286$ \\
$9$ & $0$ & $0.0104166666666666667$ & $1.37650916689918155$ & $2.06206984747023810$ \\
$10$ & $0$ & $0.00972222222222222222$ & $1.60327132694304936$ & $2.44618191189236111$ \\
$11$ & $0$ & $0.00909090909090909091$ & $1.84480803078215188$ & $2.86053214905753968$ \\
$12$ & $0$ & $0.00852272727272727273$ & $2.10064527585908964$ & $3.30488413360727814$ \\
$13$ & $0$ & $0.00801282051282051282$ & $2.37133800173880244$ & $3.77960149591619318$ \\
$14$ & $0$ & $0.00755494505494505495$ & $2.65690795708237219$ & $4.28475976834512601$ \\
$15$ & $0$ & $0.00714285714285714286$ & $2.95755558374362352$ & $4.82052962194868933$ \\
$16$ & $0$ & $0.00677083333333333333$ & $3.27335266633028923$ & $5.38700272696358817$ \\
$17$ & $0$ & $0.00643382352941176471$ & $3.60439372365816554$ & $5.98427529479518081$ \\
$18$ & $0$ & $0.00612745098039215686$ & $3.95073535413724979$ & $6.61241498537252213$ \\
$19$ & $0$ & $0.00584795321637426901$ & $4.31243096040043723$ & $7.27148133475330606$ \\
$20$ & $0$ & $0.00559210526315789474$ & $4.68951957362750529$ & $7.96152061436389144$ \\
$21$ & $0$ & $0.00535714285714285714$ & $5.08203475570206065$ & $8.68257178445300868$ \\
$22$ & $0$ & $0.00514069264069264069$ & $5.49000315882483687$ & $9.43466650646625262$ \\
$23$ & $0$ & $0.00494071146245059289$ & $5.91344725460268841$ & $10.2178313344402760$ \\
$24$ & $0$ & $0.00475543478260869565$ & $6.35238559490212554$ & $11.0320883901712198$ \\
$25$ & $0$ & $0.00458333333333333333$ & $6.80683387232599976$ & $11.8774563774913858$ \\
$26$ & $0$ & $0.00442307692307692308$ & $7.27680532931280272$ & $12.7539511441170047$ \\
$27$ & $0$ & $0.00427350427350427350$ & $7.76231127420418001$ & $13.6615862353313143$ \\
$28$ & $0$ & $0.00413359788359788360$ & $8.26336139352219191$ & $14.6003732798707909$ \\
$29$ & $0$ & $0.00400246305418719212$ & $8.77996404547459387$ & $15.5703223236731912$ \\
$30$ & $0$ & $0.00387931034482758621$ & $9.31212647332053809$ & $16.5714420860040721$ \\
$31$ & $0$ & $0.00376344086021505376$ & $9.85985498817016708$ & $17.6037401728147847$ \\
$32$ & $0$ & $0.00365423387096774194$ & $10.4231551125841751$ & $18.6672232478313144$ \\
$33$ & $0$ & $0.00355113636363636364$ & $11.0020317006137251$ & $19.7618971743249798$ \\
$34$ & $0$ & $0.00345365418894830660$ & $11.5964890355376136$ & $20.8877671314857023$ \\
$35$ & $0$ & $0.00336134453781512605$ & $12.2065309114654697$ & $22.0448377113809159$ \\
$36$ & $0$ & $0.00327380952380952381$ & $12.8321607010944887$ & $23.2331129997551454$ \\
$37$ & $0$ & $0.00319069069069069069$ & $13.4733814126370208$ & $24.4525966439230546$ \\
$38$ & $0$ & $0.00311166429587482219$ & $14.1301957377105923$ & $25.7032919099978552$ \\
$39$ & $0$ & $0.00303643724696356275$ & $14.8026060919042175$ & $26.9852017314055164$ \\
$40$ & $0$ & $0.00296474358974358974$ & $15.4906146492529964$ & $28.2983287501699751$ \\
$41$ & $0$ & $0.00289634146341463415$ & $16.1942233716816174$ & $29.6426753522061681$ \\
$42$ & $0$ & $0.00283101045296167247$ & $16.9134340342426513$ & $31.0182436976054338$ \\
$43$ & $0$ & $0.00276854928017718715$ & $17.6482482468384619$ & $32.4250357467253767$ \\
$44$ & $0$ & $0.00270877378435517970$ & $18.3986674729826899$ & $33.8630532827453423$ \\
$45$ & $0$ & $0.00265151515151515152$ & $19.1646930460624880$ & $35.3322979312341901$ \\
$46$ & $0$ & $0.00259661835748792271$ & $19.9463261834807424$ & $36.8327711771816053$ \\
$47$ & $0$ & $0.00254394079555966698$ & $20.7435679989939952$ & $38.3644743798684384$ \\
$48$ & $0$ & $0.00249335106382978723$ & $21.5564195135087909$ & $39.9274087858891520$ \\
$49$ & $0$ & $0.00244472789115646259$ & $22.3848816645565572$ & $41.5215755405887935$ \\
$50$ & $0$ & $0.00239795918367346939$ & $23.2289553146318483$ & $43.1469756981351305$ \\
\hline \hline
\end{tabular*}
\caption{Coefficients of the term including $\ln^2(z)$ in the asymptotic expansion at $z=0$.  \label{table:coeffA2}}
\end{table}

\begin{table}[htb!]
\begin{tabular*}{\textwidth}{@{\extracolsep{\fill}}*{1}{*{5}{l}}}
\hline \hline
 $j$ & $2^j \, A_{1j}^f$ & $2^j \, A_{1j}^N$ & $A_{1j}^{[1]}$ & $A_{1j}^{[8]}$ \\ 
$0$ & $0$ & $0$ & $0$ & $0$ \\
$1$ & $-0.500000000000000000$ & $0.403426409720027345$ & $0.573984868040191417$ & $2.17055845832016407$ \\
$2$ & $0.500000000000000000$ & $-0.562500000000000000$ & $1.40454734115159489$ & $1.51178083507657779$ \\
$3$ & $-0.166666666666666667$ & $0.228225469906675782$ & $0.934875138307807402$ & $0.997736024881149850$ \\
$4$ & $-0.0833333333333333333$ & $0.0637289421288980040$ & $0.775839773581947293$ & $1.14464603631361168$ \\
$5$ & $-0.0500000000000000000$ & $0.0226034437493415369$ & $1.23657194305889924$ & $1.65503482285855852$ \\
$6$ & $-0.0333333333333333333$ & $0.00832581863367395874$ & $1.40924489148011548$ & $2.04684904024043822$ \\
$7$ & $-0.0238095238095238095$ & $0.00267161275221739514$ & $1.73611866021017694$ & $2.54068805556876373$ \\
$8$ & $-0.0178571428571428571$ & $0.000326769085679754263$ & $2.01963613108457395$ & $3.04233443456408626$ \\
$9$ & $-0.0138888888888888889$ & $-0.000630760827654446418$ & $2.35121792224510425$ & $3.59825934172648065$ \\
$10$ & $-0.0111111111111111111$ & $-0.000983154457237591354$ & $2.69002495992605817$ & $4.18894355650270579$ \\
$11$ & $-0.00909090909090909091$ & $-0.00107003590722068222$ & $3.05525725734640372$ & $4.82432951234603033$ \\
$12$ & $-0.00757575757575757576$ & $-0.00104384512648363655$ & $3.43840522449766828$ & $5.50042644961007402$ \\
$13$ & $-0.00641025641025641026$ & $-0.000973776227971846659$ & $3.84370407326418780$ & $6.21950375436780425$ \\
$14$ & $-0.00549450549450549451$ & $-0.000890955694746771775$ & $4.26939652973934932$ & $6.98079796920386592$ \\
$15$ & $-0.00476190476190476190$ & $-0.000808886583311428481$ & $4.71648461383686148$ & $7.78489213223147295$ \\
$16$ & $-0.00416666666666666667$ & $-0.000732921677564811823$ & $5.18463605159559649$ & $8.63168446057531761$ \\
$17$ & $-0.00367647058823529412$ & $-0.000664706105928511902$ & $5.67412102817243384$ & $9.52136011853556861$ \\
$18$ & $-0.00326797385620915033$ & $-0.000604265845503208546$ & $6.18490239340973420$ & $10.4539408890196851$ \\
$19$ & $-0.00292397660818713450$ & $-0.000550982075137553404$ & $6.71707231979268453$ & $11.4295055517780534$ \\
$20$ & $-0.00263157894736842105$ & $-0.000504036111857871131$ & $7.27064798776431523$ & $12.4480896182831516$ \\
$21$ & $-0.00238095238095238095$ & $-0.000462603572116629509$ & $7.84567191246476886$ & $13.5097366766903596$ \\
$22$ & $-0.00216450216450216450$ & $-0.000425931162025511047$ & $8.44216604263306732$ & $14.6144764074381572$ \\
$23$ & $-0.00197628458498023715$ & $-0.000393359922042171806$ & $9.06015542998857390$ & $15.7623371145003042$ \\
$24$ & $-0.00181159420289855072$ & $-0.000364325253634250167$ & $9.69965813558890983$ & $16.9533412331771207$ \\
$25$ & $-0.00166666666666666667$ & $-0.000338347916596599626$ & $10.3606911006133588$ & $18.1875086917977147$ \\
$26$ & $-0.00153846153846153846$ & $-0.000315022440085266352$ & $11.0432681129370568$ & $19.4648562355032137$ \\
$27$ & $-0.00142450142450142450$ & $-0.000294005709272710509$ & $11.7474014260648465$ & $20.7853985059978581$ \\
$28$ & $-0.00132275132275132275$ & $-0.000275006772702577984$ & $12.4731014866934035$ & $22.1491481236013211$ \\
$29$ & $-0.00123152709359605911$ & $-0.000257778141982274278$ & $13.2203774795188210$ & $23.5561161363801837$ \\
$30$ & $-0.00114942528735632184$ & $-0.000242108529465220094$ & $13.9892373988491825$ & $25.0063122004485675$ \\
$31$ & $-0.00107526881720430108$ & $-0.000227816851115825878$ & $14.7796882888055714$ & $26.4997448187878531$ \\
$32$ & $-0.00100806451612903226$ & $-0.000214747295726806739$ & $15.5917363530702574$ & $28.0364214943387440$ \\
$33$ & $-0.000946969696969696970$ & $-0.000202765273694084681$ & $16.4253870891369883$ & $29.6163488788047327$ \\
$34$ & $-0.000891265597147950089$ & $-0.000191754083630136668$ & $17.2806453792469384$ & $31.2395328864256115$ \\
$35$ & $-0.000840336134453781513$ & $-0.000181612162237623790$ & $18.1575155769535566$ & $32.9059787946617628$ \\
$36$ & $-0.000793650793650793651$ & $-0.000172250807694287388$ & $19.0560015750359145$ & $34.6156913266240418$ \\
$37$ & $-0.000750750750750750751$ & $-0.000163592287943558989$ & $19.9761068654593792$ & $36.3686747221709129$ \\
$38$ & $-0.000711237553342816501$ & $-0.000155568262660344511$ & $20.9178345891548308$ & $38.1649327978256742$ \\
$39$ & $-0.000674763832658569501$ & $-0.000148118461678218366$ & $21.8811875791142614$ & $40.0044689982859620$ \\
$40$ & $-0.000641025641025641026$ & $-0.000141189573867644397$ & $22.8661683970448024$ & $41.8872864404790736$ \\
$41$ & $-0.000609756097560975610$ & $-0.000134734309374306502$ & $23.8727793650510899$ & $43.8133879515779675$ \\
$42$ & $-0.000580720092915214866$ & $-0.000128710605222438181$ & $24.9010225929261360$ & $45.7827761018263922$ \\
$43$ & $-0.000553709856035437431$ & $-0.000123080949939763608$ & $25.9509000018346639$ & $47.7954532330256075$ \\
$44$ & $-0.000528541226215644820$ & $-0.000117811807372872282$ & $27.0224133448867659$ & $49.8514214833131602$ \\
$45$ & $-0.000505050505050505051$ & $-0.000112873123475394251$ & $28.1155642250892192$ & $51.9506828087947926$ \\
$46$ & $-0.000483091787439613527$ & $-0.000108237902755294139$ & $29.2303541110445423$ & $54.0932390024796784$ \\
$47$ & $-0.000462534690101757632$ & $-0.000103881843409938586$ & $30.3667843507254651$ & $56.2790917109051756$ \\
$48$ & $-0.000443262411347517730$ & $-0.0000997830220740902502$ & $31.5248561835913941$ & $58.5082424487715445$ \\
$49$ & $-0.000425170068027210884$ & $-0.0000959216206473440353$ & $32.7045707512760282$ & $60.7806926118593002$ \\
$50$ & $-0.000408163265306122449$ & $-0.0000922796889249707401$ & $33.9059291070379442$ & $63.0964434884591149$ \\
\hline \hline
\end{tabular*}
\caption{Coefficients of the term including $\ln^1(z)$ in the asymptotic expansion at $z=0$..  \label{table:coeffA1}}
\end{table}

\begin{table}[htb!]
\begin{tabular*}{\textwidth}{@{\extracolsep{\fill}}*{1}{*{5}{l}}}
\hline \hline
 $j$ & $2^j \, A_{0j}^f$ & $2^j \, A_{0j}^N$ & $A_{0j}^{[1]}$ & $A_{0j}^{[8]}$ \\ 
$0$ & $0$ & $0$ & $3.42528122159600831$ & $3.42528122159600831$ \\
$1$ & $-0.704568546293369794$ & $1.46979922788286000$ & $4.66891875378567449$ & $6.16855393641111585$ \\
$2$ & $0.250000000000000000$ & $-1.18439296289188648$ & $-1.86409698775000058$ & $-2.88813944260891989$ \\
$3$ & $0.172983646604450521$ & $0.0949651522537269318$ & $-0.165543932536990844$ & $0.512133848537591203$ \\
$4$ & $-0.0454526211422191839$ & $0.230728948405102761$ & $0.783822424115675752$ & $0.859369066687737945$ \\
$5$ & $-0.0665771282408870659$ & $0.170535478708781573$ & $0.443366471441374892$ & $0.739943081737993080$ \\
$6$ & $-0.0633662336420728587$ & $0.121097475197090964$ & $0.673319154273754727$ & $0.964054976222218122$ \\
$7$ & $-0.0565427745969454660$ & $0.0876140562433534429$ & $0.726435913366923218$ & $1.11486481887254439$ \\
$8$ & $-0.0498723587254868773$ & $0.0651389700953164134$ & $0.863747347958426080$ & $1.32361727301450459$ \\
$9$ & $-0.0440616719760841497$ & $0.0497810935127544657$ & $0.976912991777519160$ & $1.53050553440405569$ \\
$10$ & $-0.0391440135067932457$ & $0.0390348840020991562$ & $1.11015101784929620$ & $1.75907755893378004$ \\
$11$ & $-0.0350044642860141860$ & $0.0313258282494950561$ & $1.24480436444187547$ & $1.99914064796198993$ \\
$12$ & $-0.0315106296098378598$ & $0.0256579376827579752$ & $1.38913091657326629$ & $2.25522618478262084$ \\
$13$ & $-0.0285450404789708972$ & $0.0213919577537874515$ & $1.53961891493419973$ & $2.52544227034301571$ \\
$14$ & $-0.0260107922038821160$ & $0.0181104142502782549$ & $1.69792429523598923$ & $2.81072538744508272$ \\
$15$ & $-0.0238299068434102881$ & $0.0155355012815971487$ & $1.86335877428942494$ & $3.11071728151804442$ \\
$16$ & $-0.0219401020018442737$ & $0.0134786933284400312$ & $2.03627829834331845$ & $3.42563011520760643$ \\
$17$ & $-0.0202916306042299759$ & $0.0118094710625241931$ & $2.21655085692990607$ & $3.75540310288813089$ \\
$18$ & $-0.0188446171160182847$ & $0.0104356584357721029$ & $2.40426173915946235$ & $4.10009310936063358$ \\
$19$ & $-0.0175669296793317974$ & $0.00929088429856948435$ & $2.59939065360775134$ & $4.45969667277959089$ \\
$20$ & $-0.0164325173540844657$ & $0.00832646471093260712$ & $2.80196284146679285$ & $4.83423421830800119$ \\
$21$ & $-0.0154201217850573928$ & $0.00750606220433089230$ & $3.01197953444744503$ & $5.22371177788899880$ \\
$22$ & $-0.0145122808533759918$ & $0.00680211295838077159$ & $3.22945112425706428$ & $5.62813964601915529$ \\
$23$ & $-0.0136945568658170378$ & $0.00619339578465035890$ & $3.45438200465618936$ & $6.04752410936692896$ \\
$24$ & $-0.0129549365005402128$ & $0.00566334979450815990$ & $3.68677804287725125$ & $6.48187173083529335$ \\
$25$ & $-0.0122833620471584281$ & $0.00519889072337732331$ & $3.92664328942136242$ & $6.93118761018387768$ \\
$26$ & $-0.0116713632118138227$ & $0.00478956469551477748$ & $4.17398173100927562$ & $7.39547643702573156$ \\
$27$ & $-0.0111117662277442632$ & $0.00442693397209045364$ & $4.42879660910536371$ & $7.87474217015394951$ \\
$28$ & $-0.0105984626639977006$ & $0.00410412467049960844$ & $4.69109086778937192$ & $8.36898833641172693$ \\
$29$ & $-0.0101262245740927213$ & $0.00381548927443140816$ & $4.96086703988589516$ & $8.87821800672572883$ \\
$30$ & $-0.00969055581419424911$ & $0.00355635166460644979$ & $5.23812738375749436$ & $9.40243390542974990$ \\
$31$ & $-0.00928757175339190598$ & $0.00332281227543720113$ & $5.52287388577884265$ & $9.94163843885498146$ \\
$32$ & $-0.00891390139930493534$ & $0.00311159761662638260$ & $5.81510831583553196$ & $10.4958337491565218$ \\
$33$ & $-0.00856660732185223946$ & $0.00291994291781584863$ & $6.11483224808581666$ & $11.0650217450417330$ \\
$34$ & $-0.00824311978897593443$ & $0.00274549977534541999$ & $6.42204709090111519$ & $11.6492041346772860$ \\
$35$ & $-0.00794118231352480694$ & $0.00258626286421113247$ & $6.73675410619610707$ & $12.2483824501994908$ \\
$36$ & $-0.00765880641186777087$ & $0.00244051132612249695$ & $7.05895442877667943$ & $12.8625580702125168$ \\
$37$ & $-0.00739423383763839559$ & $0.00230676155471644724$ & $7.38864908144816501$ & $13.4917322382236761$ \\
$38$ & $-0.00714590491210128338$ & $0.00218372890430509771$ & $7.72583898864940582$ & $14.1359060788450205$ \\
$39$ & $-0.00691243185114793832$ & $0.00207029643897324352$ & $8.07052498785055164$ & $14.7950806115583978$ \\
$40$ & $-0.00669257620669373272$ & $0.00196548927608151900$ & $8.42270783955529810$ & $15.4692567627016428$ \\
$41$ & $-0.00648522971139268382$ & $0.00186845340506846630$ & $8.78238823588775214$ & $16.1584353758174261$ \\
$42$ & $-0.00628939795077687690$ & $0.00177843810891015537$ & $9.14956680809392764$ & $16.8626172206769202$ \\
$43$ & $-0.00610418639424665989$ & $0.00169478130299057532$ & $9.52424413306341270$ & $17.5818030011429375$ \\
$44$ & $-0.00592878840194450723$ & $0.00161689724972401284$ & $9.90642073904136991$ & $18.3159933620562715$ \\
$45$ & $-0.00576247489315394670$ & $0.00154426621809640365$ & $10.2960971106333216$ & $19.0651888952766826$ \\
$46$ & $-0.00560458541710158406$ & $0.00147642574342047211$ & $10.6932736932083520$ & $19.8293901450001031$ \\
$47$ & $-0.00545452041170817043$ & $0.00141296320996856203$ & $11.0979508967802937$ & $20.6085976124493239$ \\
$48$ & $-0.00531173447211092237$ & $0.00135350953216719253$ & $11.5101290994388940$ & $21.4028117600231777$ \\
$49$ & $-0.00517573048036272032$ & $0.00129773375200797404$ & $11.9298086503898218$ & $22.2120330149751836$ \\
$50$ & $-0.00504605447193604854$ & $0.00124533840373721737$ & $12.3569898726547675$ & $23.0362617726827836$ \\
\hline \hline
\end{tabular*}
\caption{Coefficients of the term including $\ln^0(z)$ in the asymptotic expansion at $z=0$..  \label{table:coeffA0}}
\end{table}

\begin{table}[htb!]
\begin{tabular*}{\textwidth}{@{\extracolsep{\fill}}*{1}{*{5}{l}}}
\hline \hline
 $j$ & $B_{j}^f$ & $B_{j}^N$ & $B_{j}^{[1]}$ & $B_{j}^{[8]}$ \\ 
$0$ & $-0.168517703262946487$ & $0.416128621492787537$ & $3.45029758896032494$ & $3.13598994504275536$ \\
$1$ & $-0.282785299645082407$ & $0.353340865586323356$ & $0.423160291785865402$ & $2.25490716176977698$ \\
$2$ & $0.0332327652969076699$ & $-0.0789264614706424505$ & $-2.44791349327539086$ & $-1.20968260204403655$ \\
$3$ & $-0.0204682261706181858$ & $0.182572414463330030$ & $1.38064160766884538$ & $1.58944413597372890$ \\
$4$ & $-0.140378146007119793$ & $0.187648480192853526$ & $-1.31004052808255559$ & $-1.02544117114502417$ \\
$5$ & $-0.0657769561913774996$ & $0.0744359963367625279$ & $0.391266317015216009$ & $0.254747148816610245$ \\
$6$ & $-0.0768607539065821039$ & $0.0684060369878690482$ & $-1.56468633648293338$ & $-1.55664918318705640$ \\
$7$ & $-0.0469509663575984915$ & $0.0336112732298606905$ & $0.403814582313552162$ & $0.213199895820109900$ \\
$8$ & $-0.0507292429419673853$ & $0.0342481193257635669$ & $-1.53032311867262529$ & $-1.61183061501116307$ \\
$9$ & $-0.0348644177562329522$ & $0.0188492867582985753$ & $0.473639721620140887$ & $0.281523652042754200$ \\
$10$ & $-0.0369142633777180748$ & $0.0205480827739276674$ & $-1.47067196675353441$ & $-1.58329924419574094$ \\
$11$ & $-0.0271918113194230990$ & $0.0121058040405675221$ & $0.538596531943378570$ & $0.356212281904249323$ \\
$12$ & $-0.0285656719127416905$ & $0.0137358458248078444$ & $-1.41691188600314585$ & $-1.53966191681101009$ \\
$13$ & $-0.0220437744693293732$ & $0.00845821961649430542$ & $0.592231023861090150$ & $0.421556403080199613$ \\
$14$ & $-0.0230640627653818318$ & $0.00984421148015076305$ & $-1.37225043911136540$ & $-1.49666316112777735$ \\
$15$ & $-0.0184066270983388587$ & $0.00625226246459065378$ & $0.635750994218480368$ & $0.476365857313349966$ \\
$16$ & $-0.0192079008289260056$ & $0.00740595962170838745$ & $-1.33553362512433651$ & $-1.45798604315434632$ \\
$17$ & $-0.0157255741443448320$ & $0.00481274079751874196$ & $0.671271563892600617$ & $0.522147983277515918$ \\
$18$ & $-0.0163770713132275415$ & $0.00577537629588882738$ & $-1.30515039674247796$ & $-1.42408602126763659$ \\
$19$ & $-0.0136803017259498587$ & $0.00382018859773035631$ & $0.700614473187632475$ & $0.560647199219286005$ \\
$20$ & $-0.0142228895617320581$ & $0.00463052663871787744$ & $-1.27973039341601036$ & $-1.39453988844395111$ \\
$21$ & $-0.0120758381529482965$ & $0.00310648030978272061$ & $0.725174257322629249$ & $0.593340930282221275$ \\
$22$ & $-0.0125359066330851987$ & $0.00379566529823286465$ & $-1.25821292673547796$ & $-1.36873999294815108$ \\
$23$ & $-0.0107878505057337879$ & $0.00257593855214781736$ & $0.745989915897840775$ & $0.621390931213121243$ \\
$24$ & $-0.0111835123556567327$ & $0.00316804502064896411$ & $-1.23979482952551276$ & $-1.34610307629103926$ \\
$25$ & $-0.00973382477945855427$ & $0.00217074935104616770$ & $0.763835480863938651$ & $0.645693541402569071$ \\
$26$ & $-0.0100780641962329206$ & $0.00268427115089400477$ & $-1.22386774156934701$ & $-1.32612524838079132$ \\
$27$ & $-0.00885710280909751553$ & $0.00185426866015972855$ & $0.779293017237853324$ & $0.666940145862857870$ \\
$28$ & $-0.00915953604352683560$ & $0.00230348401463192284$ & $-1.20996699466727283$ & $-1.30838726883624163$ \\
$29$ & $-0.00811762725674374079$ & $0.00160234320215278568$ & $0.792805794967716454$ & $0.685667766088434601$ \\
$30$ & $-0.00838556033877359394$ & $0.00199837738781518633$ & $-1.19773366541976392$ & $-1.29254493651969908$ \\
$31$ & $-0.00748635421066146198$ & $0.00139852245186918622$ & $0.804715940749439320$ & $0.702297603675718211$ \\
$32$ & $-0.00772545404443920989$ & $0.00175013624216371005$ & $-1.18688716641738518$ & $-1.27831644326221283$ \\
$33$ & $-0.00694175785058700870$ & $0.00123128536902614487$ & $0.815291010050466256$ & $0.717163536729087055$ \\
$34$ & $-0.00715649646359657722$ & $0.00154545236922588963$ & $-1.17720553714907876$ & $-1.26547055946534249$ \\
$35$ & $-0.00646757439252508634$ & $0.00109236696069795998$ & $0.824742863220513751$ & $0.730533040526046132$ \\
$36$ & $-0.00666153314947677809$ & $0.00137469656382034773$ & $-1.16851120047889771$ & $-1.25381668226643082$ \\
$37$ & $-0.00605130517344580766$ & $0.000975712532751512072$ & $0.833241218770911649$ & $0.742622588862532182$ \\
$38$ & $-0.00622738894836428599$ & $0.00123076176163977603$ & $-1.16066057491051266$ & $-1.24319679399516649$ \\
$39$ & $-0.00568319882131712407$ & $0.000876804074393604715$ & $0.840923504935131665$ & $0.753609074006862280$ \\
$40$ & $-0.00584379031184480472$ & $0.00110830957325260412$ & $-1.15353641670660252$ & $-1.23347907651977841$ \\
$41$ & $-0.00535554391084349906$ & $0.000792214677069220274$ & $0.847902111638408179$ & $0.763638352460672414$ \\
$42$ & $-0.00550261704047495259$ & $0.00100326703462723424$ & $-1.14704211027132334$ & $-1.22455287172718359$ \\
$43$ & $-0.00506216768348881240$ & $0.000719306796472778953$ & $0.854269795614861893$ & $0.772831703715481368$ \\
$44$ & $-0.00519737237506979568$ & $0.000912482886626387005$ & $-1.14109736298269223$ & $-1.21632470597557701$ \\
$45$ & $-0.00479807448751248788$ & $0.000656023608184332253$ & $0.860103757391016805$ & $0.781290761064681086$ \\
$46$ & $-0.00492280099328466198$ & $0.000833488080047881039$ & $-1.13563492344453775$ & $-1.20871514539826241$ \\
$47$ & $-0.00455918080381595377$ & $0.000600742007224190808$ & $0.865468751501262107$ & $0.789101313296462555$ \\
$48$ & $-0.00467460917643479058$ & $0.000764325880160248162$ & $-1.13059805370458737$ & $-1.20165629743995991$ \\
$49$ & $-0.00434211822186163740$ & $0.000552167274066203553$ & $0.870419484628473632$ & $0.796336263861851858$ \\
$50$ & $-0.00444925681522386800$ & $0.000703429372660713739$ & $-1.12593856292945402$ & $-1.19508981524572365$ \\
\hline \hline
\end{tabular*}
\caption{Coefficients of  the asymptotic expansion at $z=1/2$..  \label{table:coeffB}}
\end{table}

\begin{table}[htb!]
\begin{tabular*}{\textwidth}{@{\extracolsep{\fill}}*{1}{*{5}{l}}}
\hline \hline
 $j$ & $2^j \, C_{3j}^f$ & $2^j \, C_{3j}^N$ & $C_{3j}^{[1]}$ & $C_{3j}^{[8]}$ \\ 
$0$ & $0$ & $0$ & $0$ & $0$ \\
$1$ & $0$ & $0$ & $-0.833333333333333333$ & $-1.00000000000000000$ \\
$2$ & $0$ & $0$ & $0.125000000000000000$ & $0.0625000000000000000$ \\
$3$ & $0$ & $0$ & $0$ & $0$ \\
$4$ & $0$ & $0$ & $0$ & $0$ \\
$5$ & $0$ & $0$ & $0$ & $0$ \\
$6$ & $0$ & $0$ & $0$ & $0$ \\
$7$ & $0$ & $0$ & $0$ & $0$ \\
$8$ & $0$ & $0$ & $0$ & $0$ \\
$9$ & $0$ & $0$ & $0$ & $0$ \\
$10$ & $0$ & $0$ & $0$ & $0$ \\
$11$ & $0$ & $0$ & $0$ & $0$ \\
$12$ & $0$ & $0$ & $0$ & $0$ \\
$13$ & $0$ & $0$ & $0$ & $0$ \\
$14$ & $0$ & $0$ & $0$ & $0$ \\
$15$ & $0$ & $0$ & $0$ & $0$ \\
$16$ & $0$ & $0$ & $0$ & $0$ \\
$17$ & $0$ & $0$ & $0$ & $0$ \\
$18$ & $0$ & $0$ & $0$ & $0$ \\
$19$ & $0$ & $0$ & $0$ & $0$ \\
$20$ & $0$ & $0$ & $0$ & $0$ \\
$21$ & $0$ & $0$ & $0$ & $0$ \\
$22$ & $0$ & $0$ & $0$ & $0$ \\
$23$ & $0$ & $0$ & $0$ & $0$ \\
$24$ & $0$ & $0$ & $0$ & $0$ \\
$25$ & $0$ & $0$ & $0$ & $0$ \\
$26$ & $0$ & $0$ & $0$ & $0$ \\
$27$ & $0$ & $0$ & $0$ & $0$ \\
$28$ & $0$ & $0$ & $0$ & $0$ \\
$29$ & $0$ & $0$ & $0$ & $0$ \\
$30$ & $0$ & $0$ & $0$ & $0$ \\
$31$ & $0$ & $0$ & $0$ & $0$ \\
$32$ & $0$ & $0$ & $0$ & $0$ \\
$33$ & $0$ & $0$ & $0$ & $0$ \\
$34$ & $0$ & $0$ & $0$ & $0$ \\
$35$ & $0$ & $0$ & $0$ & $0$ \\
$36$ & $0$ & $0$ & $0$ & $0$ \\
$37$ & $0$ & $0$ & $0$ & $0$ \\
$38$ & $0$ & $0$ & $0$ & $0$ \\
$39$ & $0$ & $0$ & $0$ & $0$ \\
$40$ & $0$ & $0$ & $0$ & $0$ \\
$41$ & $0$ & $0$ & $0$ & $0$ \\
$42$ & $0$ & $0$ & $0$ & $0$ \\
$43$ & $0$ & $0$ & $0$ & $0$ \\
$44$ & $0$ & $0$ & $0$ & $0$ \\
$45$ & $0$ & $0$ & $0$ & $0$ \\
$46$ & $0$ & $0$ & $0$ & $0$ \\
$47$ & $0$ & $0$ & $0$ & $0$ \\
$48$ & $0$ & $0$ & $0$ & $0$ \\
$49$ & $0$ & $0$ & $0$ & $0$ \\
$50$ & $0$ & $0$ & $0$ & $0$ \\
\hline \hline
\end{tabular*}
\caption{Coefficients of the term including $\ln^3(1-z)$ in the asymptotic expansion at $z=1$.  \label{table:coeffC3}}
\end{table}

\begin{table}[htb!]
\begin{tabular*}{\textwidth}{@{\extracolsep{\fill}}*{1}{*{5}{l}}}
\hline \hline
 $j$ & $2^j \, C_{2j}^f$ & $2^j \, C_{2j}^N$ & $C_{2j}^{[1]}$ & $C_{2j}^{[8]}$ \\ 
$0$ & $0$ & $0$ & $-0.500000000000000000$ & $-1.00000000000000000$ \\
$1$ & $0.500000000000000000$ & $0$ & $0.375000000000000000$ & $0.125000000000000000$ \\
$2$ & $0$ & $0$ & $0.500000000000000000$ & $0.156250000000000000$ \\
$3$ & $0$ & $0$ & $-0.604166666666666667$ & $-0.875000000000000000$ \\
$4$ & $0$ & $0$ & $-0.346354166666666667$ & $-0.721354166666666667$ \\
$5$ & $0$ & $0$ & $-0.350520833333333333$ & $-0.834114583333333333$ \\
$6$ & $0$ & $0$ & $-0.385026041666666667$ & $-0.973763020833333333$ \\
$7$ & $0$ & $0$ & $-0.439211309523809524$ & $-1.16114211309523810$ \\
$8$ & $0$ & $0$ & $-0.515722656250000000$ & $-1.38974260602678571$ \\
$9$ & $0$ & $0$ & $-0.613289000496031746$ & $-1.65513005332341270$ \\
$10$ & $0$ & $0$ & $-0.729432896205357143$ & $-1.95550711495535714$ \\
$11$ & $0$ & $0$ & $-0.862777925741792929$ & $-2.28958574655596140$ \\
$12$ & $0$ & $0$ & $-1.01303573101973981$ & $-2.65668519695771893$ \\
$13$ & $0$ & $0$ & $-1.17991834994703647$ & $-3.05632701361003363$ \\
$14$ & $0$ & $0$ & $-1.36316437815995311$ & $-3.48820983547126430$ \\
$15$ & $0$ & $0$ & $-1.56259955830573506$ & $-3.95211647889870546$ \\
$16$ & $0$ & $0$ & $-1.77812078629234467$ & $-4.44789395514941398$ \\
$17$ & $0$ & $0$ & $-2.00964678490501186$ & $-4.97542730649162891$ \\
$18$ & $0$ & $0$ & $-2.25711216339576779$ & $-5.53462967126235426$ \\
$19$ & $0$ & $0$ & $-2.52046641362414157$ & $-6.12543322111717852$ \\
$20$ & $0$ & $0$ & $-2.79967042411736777$ & $-6.74778438245234195$ \\
$21$ & $0$ & $0$ & $-3.09469240187353532$ & $-7.40164004055928771$ \\
$22$ & $0$ & $0$ & $-3.40550618736505688$ & $-8.08696512052357670$ \\
$23$ & $0$ & $0$ & $-3.73209019379054743$ & $-8.80373073781657916$ \\
$24$ & $0$ & $0$ & $-4.07442646337974420$ & $-9.55191288872596408$ \\
$25$ & $0$ & $0$ & $-4.43249987801023354$ & $-10.3314914466364967$ \\
$26$ & $0$ & $0$ & $-4.80629761592309479$ & $-11.1424494082543214$ \\
$27$ & $0$ & $0$ & $-5.19580873897860537$ & $-11.9847723068086583$ \\
$28$ & $0$ & $0$ & $-5.60102385956322589$ & $-12.8584477555774552$ \\
$29$ & $0$ & $0$ & $-6.02193487234525395$ & $-13.7634650861817312$ \\
$30$ & $0$ & $0$ & $-6.45853474256834786$ & $-14.6998150604822423$ \\
$31$ & $0$ & $0$ & $-6.91081733551847527$ & $-15.6674896384988558$ \\
$32$ & $0$ & $0$ & $-7.37877727719830968$ & $-16.6664817902528649$ \\
$33$ & $0$ & $0$ & $-7.86240983991090282$ & $-17.6967853419757042$ \\
$34$ & $0$ & $0$ & $-8.36171084802468743$ & $-18.7583948496213912$ \\
$35$ & $0$ & $0$ & $-8.87667659977821468$ & $-19.8513054941546909$ \\
$36$ & $0$ & $0$ & $-9.40730380197652541$ & $-20.9755129943728829$ \\
$37$ & $0$ & $0$ & $-9.95358951517831922$ & $-22.1310135339210606$ \\
$38$ & $0$ & $0$ & $-10.5155311074776748$ & $-23.3178036998784526$ \\
$39$ & $0$ & $0$ & $-11.0931262153432401$ & $-24.5358804308290445$ \\
$40$ & $0$ & $0$ & $-11.6863727102867600$ & $-25.7852409727512290$ \\
$41$ & $0$ & $0$ & $-12.2952686703745418$ & $-27.0658828413863379$ \\
$42$ & $0$ & $0$ & $-12.9198123557819007$ & $-28.3778037900026175$ \\
$43$ & $0$ & $0$ & $-13.5600021877367931$ & $-29.7210017816732323$ \\
$44$ & $0$ & $0$ & $-14.2158367303170223$ & $-31.0954749653477825$ \\
$45$ & $0$ & $0$ & $-14.8873146746601491$ & $-32.5012216551252679$ \\
$46$ & $0$ & $0$ & $-15.5744348252213637$ & $-33.9382403122397490$ \\
$47$ & $0$ & $0$ & $-16.2771960877761519$ & $-35.4065295293533892$ \\
$48$ & $0$ & $0$ & $-16.9955974589147933$ & $-36.9060880168193254$ \\
$49$ & $0$ & $0$ & $-17.7296380168167827$ & $-38.4369145906320858$ \\
$50$ & $0$ & $0$ & $-18.4793169131269618$ & $-39.9990081618285737$ \\
\hline \hline
\end{tabular*}
\caption{Coefficients of the term including $\ln^2(1-z)$ in the asymptotic expansion at $z=1$.   \label{table:coeffC2}}
\end{table}

\begin{table}[htb!]
\begin{tabular*}{\textwidth}{@{\extracolsep{\fill}}*{1}{*{5}{l}}}
\hline \hline
 $j$ & $2^j \, C_{1j}^f$ & $2^j \, C_{1j}^N$ & $C_{1j}^{[1]}$ & $C_{1j}^{[8]}$ \\ 
$0$ & $0.166666666666666667$ & $0$ & $0.373201467029786206$ & $-1.27173259981844023$ \\
$1$ & $-0.833333333333333333$ & $3.00000000000000000$ & $1.23173746150845349$ & $5.16653966205313280$ \\
$2$ & $-1.33333333333333333$ & $-1.00000000000000000$ & $-0.00494158643717262108$ & $-0.725017331411933317$ \\
$3$ & $0$ & $1.77777777777777778$ & $-0.395833333333333333$ & $-1.20486111111111111$ \\
$4$ & $0$ & $-1.16666666666666667$ & $-0.277777777777777778$ & $-0.948437500000000000$ \\
$5$ & $0$ & $1.17333333333333333$ & $-0.554513888888888889$ & $-1.41840277777777778$ \\
$6$ & $0$ & $-1.06666666666666667$ & $-0.584969618055555556$ & $-1.65718843005952381$ \\
$7$ & $0$ & $1.07755102040816327$ & $-0.760761408730158730$ & $-2.04579710530045351$ \\
$8$ & $0$ & $-1.03571428571428571$ & $-0.890788020611890590$ & $-2.41970299822402920$ \\
$9$ & $0$ & $1.04409171075837743$ & $-1.06909093878436791$ & $-2.86594082809990316$ \\
$10$ & $0$ & $-1.02222222222222222$ & $-1.26045991812523621$ & $-3.34571556929932116$ \\
$11$ & $0$ & $1.02846648301193756$ & $-1.48125274109362165$ & $-3.87678212765420378$ \\
$12$ & $0$ & $-1.01515151515151515$ & $-1.72178159994555874$ & $-4.44989914821144654$ \\
$13$ & $0$ & $1.01990317374932760$ & $-1.98624445654459479$ & $-5.06893604938213652$ \\
$14$ & $0$ & $-1.01098901098901099$ & $-2.27248555531881531$ & $-5.73156896613754091$ \\
$15$ & $0$ & $1.01470085470085470$ & $-2.58130701616566831$ & $-6.43861687526433029$ \\
$16$ & $0$ & $-1.00833333333333333$ & $-2.91208509396995555$ & $-7.18944484869064573$ \\
$17$ & $0$ & $1.01130334486735871$ & $-3.26499465151986583$ & $-7.98419148613734490$ \\
$18$ & $0$ & $-1.00653594771241830$ & $-3.63985834801095001$ & $-8.82265487551655123$ \\
$19$ & $0$ & $1.00896203356688936$ & $-4.03668457854306176$ & $-9.70483016266523527$ \\
$20$ & $0$ & $-1.00526315789473684$ & $-4.45540393460537352$ & $-10.6306360107213999$ \\
$21$ & $0$ & $1.00728010502446593$ & $-4.89599972714972990$ & $-11.6000446681061046$ \\
$22$ & $0$ & $-1.00432900432900433$ & $-5.35843853597548166$ & $-12.6130138132108190$ \\
$23$ & $0$ & $1.00603114591772437$ & $-5.84270173649283549$ & $-13.6695177291884958$ \\
$24$ & $0$ & $-1.00362318840579710$ & $-6.34876881173498351$ & $-14.7695298901327779$ \\
$25$ & $0$ & $1.00507826086956522$ & $-6.87662476770793401$ & $-15.9130301618408952$ \\
$26$ & $0$ & $-1.00307692307692308$ & $-7.42625552534829983$ & $-17.1000001086628939$ \\
$27$ & $0$ & $1.00433470507544582$ & $-7.99764952495647719$ & $-18.3304244611695035$ \\
$28$ & $0$ & $-1.00264550264550265$ & $-8.59079644181840970$ & $-19.6042897004770090$ \\
$29$ & $0$ & $1.00374333905844013$ & $-9.20568743020964121$ & $-20.9215842153437468$ \\
$30$ & $0$ & $-1.00229885057471264$ & $-9.84231466858811396$ & $-22.2822977816234047$ \\
$31$ & $0$ & $1.00326527683088737$ & $-10.5006713159098358$ & $-23.6864214614468610$ \\
$32$ & $0$ & $-1.00201612903225806$ & $-11.1807513142262801$ & $-25.1339473579526393$ \\
$33$ & $0$ & $1.00287330785864510$ & $-11.8825493052407120$ & $-26.6248684934589042$ \\
$34$ & $0$ & $-1.00178253119429590$ & $-12.6060605197156947$ & $-28.1591786673497999$ \\
$35$ & $0$ & $1.00254792826221398$ & $-13.3512807052390185$ & $-29.7368723574105435$ \\
$36$ & $0$ & $-1.00158730158730159$ & $-14.1182060551727542$ & $-31.3579446266081720$ \\
$37$ & $0$ & $1.00227486173432119$ & $-14.9068331534909395$ & $-33.0223910493489249$ \\
$38$ & $0$ & $-1.00142247510668563$ & $-15.7171589252681422$ & $-34.7302076462865889$ \\
$39$ & $0$ & $1.00204346358192512$ & $-16.5491805956531018$ & $-36.4813908299808055$ \\
$40$ & $0$ & $-1.00128205128205128$ & $-17.4028956540331996$ & $-38.2759373577121808$ \\
$41$ & $0$ & $1.00184566573620708$ & $-18.2783018234886591$ & $-40.1138442912291208$ \\
$42$ & $0$ & $-1.00116144018583043$ & $-19.1753970342317128$ & $-41.9951089618498173$ \\
$43$ & $0$ & $1.00167526283158992$ & $-20.0941794006528438$ & $-43.9197289403202292$ \\
$44$ & $0$ & $-1.00105708245243129$ & $-21.0346472013101126$ & $-45.8877020105841763$ \\
$45$ & $0$ & $1.00152741889175998$ & $-21.9967988614846210$ & $-47.8990261469401197$ \\
$46$ & $0$ & $-1.00096618357487923$ & $-22.9806329379004151$ & $-49.9536994940594338$ \\
$47$ & $0$ & $1.00139832000402394$ & $-23.9861481053159980$ & $-52.0517203494706312$ \\
$48$ & $0$ & $-1.00088652482269504$ & $-25.0133431447191223$ & $-54.1930871481555025$ \\
$49$ & $0$ & $1.00128492560723812$ & $-26.0622169329093418$ & $-56.3777984489697630$ \\
$50$ & $0$ & $-1.00081632653061224$ & $-27.1327684332803054$ & $-58.6058529226396407$ \\
\hline \hline
\end{tabular*}
\caption{Coefficients of the term including $\ln^1(1-z)$ in the asymptotic expansion at $z=1$.   \label{table:coeffC1}}
\end{table}

\begin{table}[htb!]
\begin{tabular*}{\textwidth}{@{\extracolsep{\fill}}*{1}{*{5}{l}}}
\hline \hline
 $j$ & $2^j \, C_{0j}^f$ & $2^j \, C_{0j}^N$ & $C_{0j}^{[1]}$ & $C_{0j}^{[8]}$ \\ 
$0$ & $-0.500000000000000000$ & $1.46314842564486910$ & $2.99088036610520918$ & $5.36723418130007662$ \\
$1$ & $-1.42995604456548429$ & $1.13786367772956345$ & $1.54136149477679947$ & $2.69552794843987522$ \\
$2$ & $1.91666666666666667$ & $-2.25000000000000000$ & $1.71840214836119185$ & $-1.60075191670299076$ \\
$3$ & $0.611111111111111111$ & $-0.619590197523060191$ & $-2.06424405679334403$ & $-1.54274837758763161$ \\
$4$ & $-0.643518518518518519$ & $0.195957918976360102$ & $-0.0130413759325362400$ & $-0.303418671732870501$ \\
$5$ & $0.340277777777777778$ & $-0.0413739748096641703$ & $-0.238311320438480490$ & $-0.669957616964990730$ \\
$6$ & $-0.325555555555555556$ & $0.0506324253921429927$ & $-0.295055645061050557$ & $-0.764832722561696636$ \\
$7$ & $0.226455026455026455$ & $-0.00851301037571829193$ & $-0.339822377969461626$ & $-0.890888666312229293$ \\
$8$ & $-0.218395691609977324$ & $0.0213645648860381302$ & $-0.390146640251603639$ & $-1.05120169323700895$ \\
$9$ & $0.169229497354497354$ & $-0.00211637936795479708$ & $-0.461322294631967458$ & $-1.22313790527806617$ \\
$10$ & $-0.164403292181069959$ & $0.0113522482562368827$ & $-0.541819005550924529$ & $-1.41401698802520976$ \\
$11$ & $0.135016835016835017$ & $-0.000318767702387816365$ & $-0.628652981032540540$ & $-1.61931303433249689$ \\
$12$ & $-0.131841138659320478$ & $0.00690440469013820865$ & $-0.723275610559353758$ & $-1.84068889574422410$ \\
$13$ & $0.112292799792799793$ & $0.000265248682886861854$ & $-0.825993649701121225$ & $-2.07711487642799655$ \\
$14$ & $-0.110054475439090824$ & $0.00458868022511893992$ & $-0.936463419830444517$ & $-2.32894205920429107$ \\
$15$ & $0.0961102389673818245$ & $0.000453675356203840857$ & $-1.05451445367967720$ & $-2.59591507987482642$ \\
$16$ & $-0.0944510582010582011$ & $0.00324662943305108117$ & $-1.18019754613078527$ & $-2.87809685278666450$ \\
$17$ & $0.0840022467320261438$ & $0.000497598372753204857$ & $-1.31350469855437532$ & $-3.17541409467291957$ \\
$18$ & $-0.0827245931052159426$ & $0.00240631119711725349$ & $-1.45441091407325254$ & $-3.48786946260442032$ \\
$19$ & $0.0746028105849227280$ & $0.000486324611017114811$ & $-1.60289716755944297$ & $-3.81543598200621431$ \\
$20$ & $-0.0735893095311377860$ & $0.00184846951985198255$ & $-1.75895767015312900$ & $-4.15810594606866216$ \\
$21$ & $0.0670948203842940685$ & $0.000454723184885487464$ & $-1.92258463118975708$ & $-4.51586599684942768$ \\
$22$ & $-0.0662715591287019858$ & $0.00146081814939817257$ & $-2.09377083340939026$ & $-4.88870833600777270$ \\
$23$ & $0.0609596815130411968$ & $0.000417153244503838785$ & $-2.27251009127380187$ & $-5.27662469184229497$ \\
$24$ & $-0.0602778732504630425$ & $0.00118135287734891089$ & $-2.45879765495347611$ & $-5.67960881210003117$ \\
$25$ & $0.0558524557165861514$ & $0.000379613169515239221$ & $-2.65262925354447558$ & $-6.09765492333051935$ \\
$26$ & $-0.0552786324786324786$ & $0.000973714753442926804$ & $-2.85400116252291297$ & $-6.53075822577325613$ \\
$27$ & $0.0515348089707064066$ & $0.000344508854685172933$ & $-3.06291012971859825$ & $-6.97891445117602013$ \\
$28$ & $-0.0510452624650155514$ & $0.000815510523262436645$ & $-3.27935333968229836$ & $-7.44211991713040979$ \\
$29$ & $0.0478367911660194090$ & $0.000312657164317843491$ & $-3.50332830748461698$ & $-7.92037136980218281$ \\
$30$ & $-0.0474142649529094238$ & $0.000692373054149202175$ & $-3.73483283977384136$ & $-8.41366595367185046$ \\
$31$ & $0.0446339554237218308$ & $0.000284170609941448671$ & $-3.97386499340833642$ & $-8.92200113798354692$ \\
$32$ & $-0.0442655942883570355$ & $0.000594760509978210608$ & $-4.22042304322402568$ & $-9.44537467932493558$ \\
$33$ & $0.0418330864084392310$ & $0.000258861306961717211$ & $-4.47450545037307560$ & $-9.98378457860793908$ \\
$34$ & $-0.0415091197176758674$ & $0.000516145276015864326$ & $-4.73611083846747987$ & $-10.5372290506020856$ \\
$35$ & $0.0393629822695566640$ & $0.000236428584055178688$ & $-5.00523797276876462$ & $-11.1057064953503955$ \\
$36$ & $-0.0390758525632475212$ & $0.000451945718064507804$ & $-5.28188574264949978$ & $-11.6892154752733761$ \\
$37$ & $0.0371683191127635572$ & $0.000216545822431606155$ & $-5.56605314633079196$ & $-12.2877546949563884$ \\
$38$ & $-0.0369120898992876233$ & $0.000398873057149122539$ & $-5.85773927795625419$ & $-12.9013229841730655$ \\
$39$ & $0.0352054590281737096$ & $0.000198899272480215157$ & $-6.15694331642408975$ & $-13.5299192831255451$ \\
$40$ & $-0.0349754028397753094$ & $0.000354519337003612221$ & $-6.46366451574882714$ & $-14.1735426297905065$ \\
$41$ & $0.0334395194913487596$ & $0.000183203748098786725$ & $-6.77790219668855110$ & $-14.8321921489235718$ \\
$42$ & $-0.0332318253496123811$ & $0.000317090217589713661$ & $-7.09965573947222964$ & $-15.5058670425277342$ \\
$43$ & $0.0318422825107858736$ & $0.000169207253098887708$ & $-7.42892457745216511$ & $-16.1945665815381154$ \\
$44$ & $-0.0316538443228545987$ & $0.000285227395508213713$ & $-7.76570819155039469$ & $-16.8982900985592144$ \\
$45$ & $0.0303906756073775101$ & $0.000156690471765192253$ & $-8.11000610538431988$ & $-17.6170369814965810$ \\
$46$ & $-0.0302189365474389629$ & $0.000257887979148686607$ & $-8.46181788097831732$ & $-18.3508066679605181$ \\
$47$ & $0.0290656501330997570$ & $0.000145464061301787852$ & $-8.82114311498037979$ & $-19.0995986403335494$ \\
$48$ & $-0.0289084864727738741$ & $0.000234260924936777351$ & $-9.18798143531612041$ & $-19.8634124214126006$ \\
$49$ & $0.0278513407431530532$ & $0.000135365190944127144$ & $-9.56233249822233808$ & $-20.6422475705490873$ \\
$50$ & $-0.0277069739460410014$ & $0.000213708110211028936$ & $-9.94419598561106767$ & $-21.4361036802221302$ \\
\hline \hline
\end{tabular*}
\caption{Coefficients of the term including $\ln^0(1-z)$ in the asymptotic expansion at $z=1$.   \label{table:coeffC0}}
\end{table}

\providecommand{\href}[2]{#2}\begingroup\raggedright\endgroup


\end{document}